\documentclass[10pt,twocolumn]{IEEEtran}
\usepackage{mathrsfs}
\usepackage{amsthm}
\usepackage{amssymb}
\usepackage{graphicx}
\usepackage{multirow}
\usepackage[cmex10]{amsmath}
\usepackage{subfig}

\newcommand{\abs}[1]{|#1|}

\newcommand{\defn}{\triangleq}
\newcommand{\eu}[1]{\left\|{#1}\right\|_{2}}

\newtheorem{theorem}{Theorem}
\newtheorem{lemma}{Lemma}

\newtheorem{define}{Definition}

\newtheorem{proposition}{Proposition}
\usepackage{subeqnarray}
\usepackage{cases}
\allowdisplaybreaks
\usepackage{changebar}
\usepackage{color}
\begin{document}

\title{Detecting Byzantine Attacks Without Clean Reference}

\author{Ruohan Cao, Tan
  F. Wong, Tiejun Lv,~Hui Gao and~Shaoshi~Yang

\thanks{Manuscript received February 1, 2016; revised June 21, 2016; Accepted July 8, 2016.
The associate editor coordinating the review of the manuscript and approving
it for publication was Prof. Lifeng Lai. }
\thanks{The financial support of the National Natural
    Science Foundation of China (NSFC) (Grant No. 61271188, 61401041
    and 61501046), of the Project funded by China Postdoctoral Science Foundation,
    and of the National Science
    Foundation (NSF) under Grant CCF-1320086 is gratefully acknowledged.
}
\thanks{Copyright (c) 2016 IEEE. Personal use of this material is permitted. However, permission to use this material for any other purposes must be obtained from the IEEE by sending a request to pubs-permissions@ieee.org.
}
    \thanks{R.~Cao is with the
    Institute of Information Photonics and Optical Communications,
    Beijing University of Posts and Telecommunications (BUPT), Beijing
    100876, China (e-mail: caoruohan@bupt.edu.cn). She is also with
    the School of Information and Communication Engineering, BUPT. }
  \thanks{T. F. Wong is with the Department of Electrical and Computer
    Engineering, University of Florida, FL 32611, U.S. (e-mail:
    twong@ufl.edu).}  \thanks{T.~Lv and H.~Gao are with the School of
    Information and Communication Engineering, BUPT, Beijing 100876,
    China (e-mail: \{lvtiejun, huigao\}@bupt.edu.cn).}
  \thanks{S. Yang is with the School of Electronics and Computer
    Science, University of Southampton, SO17 1BJ Southampton,
    U.K. (e-mail: sy7g09@ecs.soton.ac.uk).}


}

\markboth{Accepted to appear on IEEE Transactions on Information Forensics and Security, Jul. 2016}%
{Shell \MakeLowercase{\textit{et al.}}: Bare Demo of IEEEtran.cls
for Journals}

\maketitle

\begin{abstract}
  We consider an amplify-and-forward relay network composed of a
  source, two relays, and a destination. In this network, the two
  relays are untrusted in the sense that they may perform Byzantine
  attacks by forwarding altered symbols to the destination. Note that
  every symbol received by the destination may be altered, and hence
  no clean reference observation is available to the destination. For
  this network, we identify a large family of Byzantine attacks that can be
  detected in the physical layer. We further investigate how the
  channel conditions impact the detection against this family of 
  attacks. In particular, we prove that all Byzantine attacks in this family
  can be detected with asymptotically small miss detection and false
  alarm probabilities by using a sufficiently large number of channel
  observations \emph{if and only if} the network satisfies a
  non-manipulability condition.  No pre-shared secret or secret
  transmission is needed for the detection of these attacks,
  demonstrating the value of this physical-layer security technique
  for counteracting Byzantine attacks.
\end{abstract}
\IEEEpeerreviewmaketitle
\section{Introduction}
\label{sec:intro}

In many communication networks, no direct physical link exists between
source and destination nodes. As a result, the information that is to
be delivered from a source to a destination has to be relayed by
intermediate relay nodes. This gives the potentially malicious relay
nodes the chances to perform Byzantine attacks by altering the
information intended for the destination.  Such attacks degrade the
security of the network, and may diminish the potential benefits of
relaying in practice \cite{Buttyan2006Security, bloch2011physical}.

Under the risk of Byzantine attacks, one major challenge in achieving
secure communications is to determine the existence of malicious
relays. Conventionally, cryptographic keys or tracing symbols are
often added to the information in or above the physical layer for
making Byzantine attacks detectable.  More specifically, in
\cite{papadimitratos2006secure}, \cite{hu2005ariadne}, cryptographic
keys are applied at the source to encrypt the data above the physical
layer.  Relying on the \textit{a priori} shared knowledge of
cryptographic keys, the destination performs attack detection by
checking whether or not its received data obeys the constraints
imposed by the cryptographic keys.
{In~\cite{mao2007tracing}-\cite{nonherentsCL},
    tracing symbols are inserted into the transmitted signal in the
    physical layer. Using these tracing symbols, the intended
    destination is capable of determining whether an attack has
    occurred or not by comparing the known and observed tracing
    symbols.
    These cryptography-based and tracing-symbol-based schemes require
    the cryptographic information and tracing symbols, to which the
    relays are not privy, to be shared between the source and
    destination before the communication takes place.}

It is also possible to detect Byzantine attacks without assuming any
prior shared secret in certain network scenarios,
where a ``clean'' reference is available for attack detection. For
instance, in \cite{ho2008byzantine}--\cite{JaggiTIT}, special symbols
are carefully generated and inserted into
network-coded packets. By comparing the 
network-coded packets received from different relays, the destination
can infer whether the relay network is malicious or not. However,
the detection methods employed by
\cite{ho2008byzantine}--\cite{JaggiTIT} require at least one guaranteed
``safe packet'' be delivered to the destination so that this safe packet
can serve as a clean reference to enable attack detection. This
requirement is often satisfied by assuming that at least one relay or
link is absolutely trustworthy.  This assumption can be 
eliminated in a two-way relay
(TWR) or one-way relay network by the schemes in \cite{he2013strong}--\cite{KimTWC}.
Benefiting from the network topology of a TWR network, each source
(destination) node can utilize its own transmitted symbols as the
clean reference to carry out attack detection \cite{he2013strong}--\cite{GravesISIT13}. To further elaborate,
in \cite{he2013strong}, each node's own lattice-coded transmitted
symbols are employed to simultaneously support secret transmission and
construct an algebraic manipulation detection (AMD) code to detect
Byzantine attacks in TWR networks with Gaussian channels. However, it
is difficult to extend this scheme to non-Gaussian channels. In our
previous work \cite{GravesINFOCOM12}--\cite{GravesISIT13}, we show
that for TWR networks with discrete memoryless channels (DMCs), it is
possible to detect potential Byzantine attacks without any AMD code or
cryptographic keys. The basic idea is that each node utilizes its own
transmitted symbols as a clean reference for statistically checking
against the other node's symbols forwarded by the
relay. 
{In \cite{KimTWC}, a two-hop, one-way
    relay network is considered. It is assumed that a safe direct link
    from the source to the destination exists. Since the direct link
    is not attacked, observations made through the direct link can be
    utilized as the clean reference to check against attacks imposed by
    the possibly malicious relay.}

In many relay networks, the clean reference assumed in
\cite{he2013strong}--\cite{KimTWC} does not exist.  Schemes have been
proposed in \cite{cao2013detecting}--\cite{OFDM} to detect Byzantine
attacks without using any prior shared secret in some relay
networks where no clean reference is available to the
destination. These schemes assume that each malicious relay garbles
its received symbols according to independent and identically
distributed (i.i.d.) stochastic distributions.  This model of
i.i.d. attacks may not always be valid in practice, although it makes
analysis simple.  The Byzantine attack detection methods presented in
\cite{cao2013detecting}--\cite{OFDM} may no longer be provably
unbreakable for non-i.i.d. attacks.

Against this background, we consider in this paper the Byzantine
attack detection problem in a one-way, two-hop network consisting of a
source node, two potentially malicious relay nodes and a destination
node, where clean reference is unavailable. There are two independent transmission paths from the source
node to the destination node, each via a relay node that may perform
Byzantine attacks. For this network, we propose an attack detection
approach employing only the physical-layer signals that the
destination receives. No AMD codes or cryptographic keys are
needed. Our treatment is considerably different from existing
contributions \cite{cao2013detecting}--\cite{OFDM} in that a more
generalized attack model is considered. In particular, this model
allows the two relays to conduct non-i.i.d. attacks. It is worth
noting that considering non-i.i.d. attacks significantly complicates
the analysis presented in this paper. Moreover, since all symbols
observed by the destination node are prone to Byzantine attacks, no
clean reference is available at the destination. Under these
more general assumptions, the major contributions of this paper are
summarized as follows:
\begin{enumerate}

\item We identify a large family of Byzantine attacks that physically
  correspond to the case in which the two relays do not collude in
  attack. For a Byzantine attack in this family, the attack can be
  sufficiently characterized by two stochastic matrices, each
  containing the empirical transition probabilities between a relay's
  input symbols and output symbols. This result is summarized in
  Proposition~1 of Section~III.


\item We prove that under a non-manipulable channel condition, all
  Byzantine attacks in the aforementioned family are asymptotically
  detectable by simply comparing proper statistics generated from the
  destination's observations to known values of the statistics when
  there is no attack. This result is summarized in Theorem~1 of 
  Section~III.

\item Additionally, we also find a non-i.i.d. attack strategy, not
  belonging to the aforementioned family of Byzantine attacks, which is
  always capable of fooling the destination in this network with no
  clean reference. This result is summarized in Proposition~2 of Section~III.
\end{enumerate}

In short, we investigate in this paper both the feasibility and
vulnerability of detecting Byzantine attacks in the physical layer of
networks without clean references.  The rest of this paper is
organized follows. In Section~II, the problem to be addressed is
formalized. In Section~III, we provide our main results as described above.
In Section~IV, we discuss how to check for the above-mentioned
non-manipulability condition numerically. Numerical simulation results
are given in Section~V, and finally the conclusions are drawn in
Section~VI. In Appendices~A and~B, we detail the proofs of
the propositions and Theorem~1, respectively.
\begin{figure}
\centering
\includegraphics[width=0.48\textwidth]{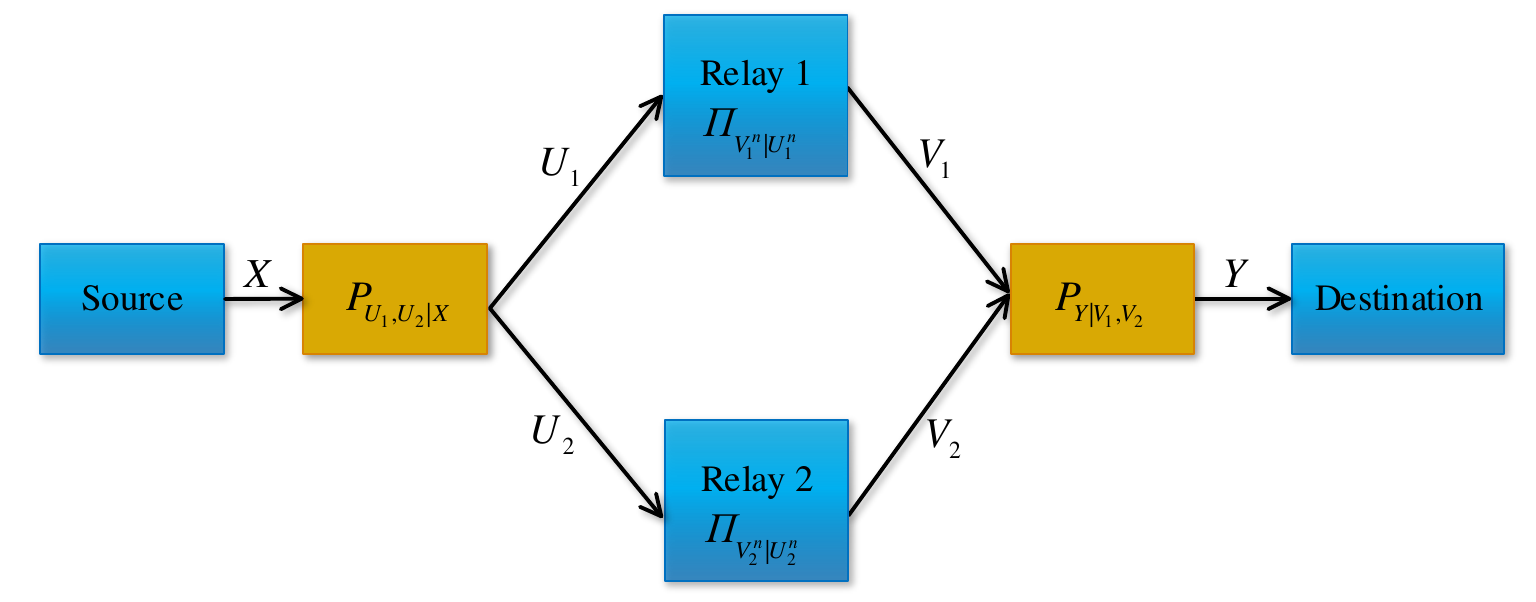}
\caption{A relay network with two unsafe links from the source to the
  destination.
}
\label{fig:model}
\end{figure}

\section{ System Model} \label{se:model}

\subsection{Notation}
Let $\bold{a}$ be an $m\!\times\!1$ column vector and $A$ be an
$m\!\times\! n$ matrix. For $i=1,2,\ldots,m$ and $j=1,2,\ldots,n$,
$[\bold{a}]_i$ denotes the $i$th element of $\bold{a}$, and
$[A]_{i,j}$ denotes the $(i,j)$th entry of $A$. Whenever there is no
ambiguity, we will employ the notation with no brackets for
simplicity.
The transpose of $A$ is denoted by $A^T$.
The Euclidean norm of $\bold{a}$ is $\left\Vert \bold{a}\right\Vert_2$,
and $\left\Vert {A}\right\Vert_2$ is the Frobenius norm of
$A$. $\abs{\bold{a}}$ and $\abs{A}$ denote the $L_1$-norm of
$\bold{a}$ and $A$, respectively.  The Kronecker product of matrices
$A$ and $B$ is represented by $A\otimes B$.  The identity and zero
matrices of any dimension are denoted by the generic symbols $I$ and
$0$, respectively.

For the random variables, we use upper-case script letters and
serif-font letters to denote the corresponding discrete alphabets and
elements in the alphabets, respectively.  For instance, suppose that
we denote a finite alphabet by
$\mathcal{X} = \{ \mathsf{x}_1, \mathsf{x}_2, \ldots,
\mathsf{x}_{|\mathcal{X}|} \}$,
where $\abs{\mathcal{X}}$ is the cardinality of $\mathcal{X}$. Then
$X$ is a generic random variable over $\mathcal{X}$, and $\mathsf{x}$
is a generic element in $\mathcal{X}$.
For a pair of random variables $X$ and $Y$, we use $P_{X}(\mathsf{x})$
and $P_{X|Y}( \mathsf{x}| \mathsf{y})$ to denote the marginal
distribution of $X$ and the conditional distribution of $X$ given $Y$,
respectively. Whenever needed, we will treat $P_{X|Y}$ as a stochastic
matrix whose entries are specified by
$P_{X|Y}( \mathsf{x}| \mathsf{y})$ with $\mathsf{x}$ and $\mathsf{y}$
indexing the rows and columns, respectively. Similarly, $P_{X}$ is
also treated as an $1\times |\mathcal{X}|$ vector whose entries are
specified by $P_{X}( \mathsf{x})$ with $\mathsf{x}$ indexing the
columns.

We employ $x^{n}$ to denote a sequence of $n$ symbols drawn from
$\mathcal{X}$, and $x_i$ to denote the $i$th symbol in $x^n$. We also
employ $X^n$ to denote a sequence of $n$ random variables defined over
$\mathcal{X}$, and $X_i$ to denote the $i$th random variable in
$X^n$. The counting function $N(\mathsf{x}|x^n)$ records the number of
occurrences of the element $\mathsf{x}$ in the sequence $x^n$. The
indicator function $1_i(\mathsf{x}|x^n)$ tells whether $x_i$ is
$\mathsf{x}$. We may also use $1_i(\mathsf{x})$ instead of
$1_i(\mathsf{x}|x^n)$ for simplicity, whenever the meaning is clear
from the context. It is clear that
$N(\mathsf{x}|x^n) = \sum_{i=1}^{n} 1_i(\mathsf{x}|x^n)$. We may
trivially extend the aforementioned notations to a tuple of symbols
drawn from the corresponding alphabets.  The empirical distribution of
the sequence $x^n$ is denoted by
$\varPi_{x^{n}}(\mathsf{x}) \triangleq \frac{1}{n}
N(\mathsf{x}|x^{n})$,
while the empirical conditional distribution of $x^{n}$ given $y^{n}$
is
$\varPi_{x^{n}|y^{n}}(\mathsf{x} | \mathsf{y}) \triangleq
\frac{N(\mathsf{x},\mathsf{y}|x^{n},y^{n})}{N(\mathsf{y}|y^{n})}$,
provided that $N(\mathsf{y}|y^{n}) >0$. Similar to the notation of
$P_{X|Y}$ and $P_{X|Y}( \mathsf{x}| \mathsf{y})$, we also treat
$\varPi_{x^{n}|y^{n}}$ as a stochastic matrix, whose entries are
specified by $\varPi_{x^{n}|y^{n}}( \mathsf{x}| \mathsf{y})$ with
$\mathsf{x}$ and $\mathsf{y}$ indexing the rows and columns,
respectively. Moreover, since $X^{n}$ and $Y^{n}$ denote random
sequences, if the subscript of $\varPi_{x^{n}|y^{n}}$ is written in
uppercase as $\varPi_{X^{n}|Y^{n}}$, then $\varPi_{X^{n}|Y^{n}}$ will
be viewed as a random matrix denoting the empirical conditional
distribution of $X^{n}$ given $Y^{n}$.

Throughout this paper, Greek letters are exclusively used to represent
constants and functions that have arbitrarily small yet positive
value.  In particular, $\varepsilon\left(\sigma\right) \rightarrow 0$
as $\sigma\rightarrow0$.  Unless otherwise stated, any convergence
involving random variables should be interpreted as convergence in
probability. For example, if $\{X^n\}$ and $\{Y^n\}$ are two sequences
of random variables (matrices) over the same alphabet,
$X^n \rightarrow Y^n$ means $|X^n - Y^n| \rightarrow 0$
(or $\|X^n - Y^n\|_2 \rightarrow 0$) in probability as $n$ approaches
infinity.  For easy reference, the main quantities defined and employed in
the rest of the paper are listed in Table I.

{\begin{table*}[!t]
\centering
\caption{Notation Table}
\begin{tabular}{|c|c|}
\hline
  $X$&
 	\text{ Source symbol}\\

 $U_{m}$&
 	\text{Symbol received by relay $\ensuremath{m\,\left(m=1,2\right)}$}\\

 $V_{m}$&
 	\text{Symbol forwarded by relay \ensuremath{m\,\left(m=1,2\right)}}\\

$Y$&
 	\text{Symbol observed by the destination}\\
	
$ \mathcal{X},\:\mathcal{U}_{m},\:\mathcal{V}_{m},\:\mathcal{Y}$ &
	Alphabets of $X$, $U_{m}$, $V_{m}$ and $Y$, respectively\\
$P_{U_{1},U_{2}|X}$&
 	\text{Conditional PMF of  \ensuremath{(U_{1},U_{2})} given \ensuremath{X}}\\
 $P_{U_{1},U_{2}}$ &
 	\text{Joint PMF of  \ensuremath{(U_{1},U_{2})}}\\
 $P_{Y|V_{1},V_{2}}$ &
 	\text{Conditional PMF of \ensuremath{Y} given
                       \ensuremath{(V_{1},V_{2})}}\\
 $\left(P_{U_{1},U_{2}},P_{Y|V_{1},V_{2}}\right)$&
	Observation channel \\
%
 $U_{m}^{n}$ &
 	\text{Symbol sequence received  by relay \ensuremath{m\,\left(m=1,2\right)}}\\
%
%
 $V_{m}^{n}$ &
 	\text{Symbol sequence forwarded by relay \ensuremath{m\,\left(m=1,2\right)}}\\
 $Y^{n}$ &
	 Symbol sequence observed by the destination \\
%
%
	
$U_{m,i}$, $V_{m,i}$ &
	The $i$th symbols in $U_{m}^{n}$ and $V_{m}^{n}$, respectively\\

$\mathsf{u}_{m}$, $\mathsf{v}_{m}$ &
	Generic elements in $\mathcal{U}_{m}$ and $\mathcal{V}_{m}$, respectively\\

$1_{i}\left(\mathsf{u}_{m}\left|U_{m}^{n}\right.\right)$ &
	Indicator of whether $U_{m,i}=\mathsf{u}_{m}$\\

$1_{i}\left(\mathsf{u}_{1}, \mathsf{u}_{2}\left|U_{1}^{n}, U_{2}^{n}\right.\right)$ &
	           Indicator of whether $U_{1,i}=\mathsf{u}_{1}$ and $U_{2,i}=\mathsf{u}_{2}$\\

$1_{i}\left(\mathsf{u}_{m},\mathsf{v}_{m}\left|U_{m}^{n},V_{m}^{n}\right.\right)$ &
	Indicator of whether $U_{m,i}=\mathsf{u}_{m}$ and $V_{m,i}=\mathsf{v}_{m}$\\

$N\left(\mathsf{u}_{m}\left|U_{m}^{n}\right.\right)$&
	$\sum_{i}^{n}1_{i}\left(\mathsf{u}_{m}\left|U_{m}^{n}\right.\right)$\\

$N\left(\mathsf{u}_{1}, \mathsf{u}_{2}\left|U_{1}^{n}, U_{2}^{n}\right.\right)$&
	$\sum_{i}^{n}1_{i}\left(\mathsf{u}_{1}, \mathsf{u}_{1} \left|U_{1}^{n}, U_{2}^{n}\right.\right)$\\

$N\left(\mathsf{u}_{m},\mathsf{v}_{m}\left|U_{m}^{n},V_{m}^{n}\right.\right)$&
	$\sum_{i}^{n}1_{i}\left(\mathsf{u}_{m},\mathsf{v}_{m}\left|U_{m}^{n},V_{m}^{n}\right.\right)$\\

$\varPi_{V_{1}^{n}\left|U_{1}^{n}\right.}\left(\mathsf{v}_{1}\left|\mathsf{u}_{1}\right.\right)$ &
	$\frac{N\left(\mathsf{u}_{1},\mathsf{v}_{1}\left|U_{1}^{n},V_{1}^{n}\right.\right)}{N\left(\mathsf{u}_{1}\left|U_{1}^{n}\right.\right)}$\\

$\varPi_{V_{2}^{n}\left|U_{2}^{n}\right.}\left(\mathsf{v}_{2}\left|\mathsf{u}_{2}\right.\right)$ &
	$\frac{N\left(\mathsf{u}_{2},\mathsf{v}_{2}\left|U_{2}^{n},V_{2}^{n}\right.\right)}{N\left(\mathsf{u}_{2}\left|U_{2}^{n}\right.\right)}$\\	
$\varPi_{V_{1}^{n},V_{2}^{n}\left|U_{1}^{n},U_{2}^{n}\right.}\left(\mathsf{v}_{1},\mathsf{v}_{2}\left|\mathsf{u}_{1},\mathsf{u}_{2}\right.\right)$&
	$\frac{N\left(\mathsf{u}_{1},\mathsf{u}_{2},\mathsf{v}_{1},\mathsf{v}_{2}\left|U_{1}^{n},U_{2}^{n},V_{1}^{n},V_{2}^{n}\right.\right)}{N\left(\mathsf{u}_{1},\mathsf{u}_{2}\left|U_{1}^{n},U_{2}^{n}\right.\right)}$\\

$\mathsf{x}_{i}$, $\mathsf{u}_{m,i}$, $\mathsf{v}_{m,i}$, $\mathsf{y}_{i}$ &
	The $i$th elements in alphabets $\mathcal{X}$, $\mathcal{U}_{m}$, $\mathcal{V}_{m}$, and $\mathcal{Y}$, respectively\\

$\left[\varPi_{V_{m}^{n}\left|U_{m}^{n}\right.}\right]_{i,j}$&
 $\varPi_{V_{m}^{n}\left|U_{m}^{n}\right.}\left(\mathsf{v}_{m,i}\left|\mathsf{u}_{m,j}\right.\right)$\\
 
$
  \left[\varPi_{V_{1}^{n},V_{2}^{n}\left|U_{1}^{n},U_{2}^{n}\right.}\right]_{i,j}$ &
 {\small{$\varPi_{V_{1}^{n},V_{2}^{n}\left|U_{1}^{n},U_{2}^{n}\right.}\left(\mathsf{v}_{1,t_{1}},\mathsf{v}_{2,t_{2}}\left|
 \mathsf{u}_{1,k_{1}}, \mathsf{u}_{2,k_{2}}\right.\right)$, 
$j=\left(k_{1}-1\right)\left|\mathcal{U}_{2}\right|+k_{2},
 i=\left(t_{1}-1\right)\left|\mathcal{U}_{2}\right|+t_{2}$}}\\

 $[P_{U_{1},U_{2}}]_{j}$ &
    $\Pr\{U_1=\mathsf{u}_{1,k}, U_2=\mathsf{u}_{2,t}\}$, $j=\left(k-1\right)\left|\mathcal{U}_{2}\right|+t$\\
 
$\left[P_{Y\left|V_{1},V_{2}\right.}\right]_{i,j}$ &
     $\Pr\{Y=\mathsf{y}_{i} \,|\, V_1=\mathsf{v}_{1,k},
                                                     V_2=\mathsf{v}_{2,t}\}$,
                                                     $j=\left(k-1\right)\left|\mathcal{U}_{2}\right|+t$\\
 $D\left(Y^{n}\right)$ &
	Decision statistic as a function of $Y^{n}$\\
 $I$ &
	Identity matrix with any dimension \\
$\otimes$ &
	Kronecker operator \\

\hline
\end{tabular}
\end{table*}}


\subsection{Channel Model} \label{sec:model}

Consider the relay network shown in Fig.~\ref{fig:model}. The source
sends symbols to the destination through two distinct paths. Along
each path, the source symbols are forwarded by a potentially malicious
relay to the destination.  A malicious relay may forward symbols that
are different from the ones received from the source.  Our goal is to
detect malicious actions of the two relays by observing the received
symbols. The channels from the source to the relays, and from the
relays to the destination are assumed to be memoryless with discrete inputs and outputs.

Let $X$ be a discrete random variable, with probability mass function (PMF) $P_X$, that specifies
a generic symbol transmitted by the source. During the period spanning
the time instants $1,2,\ldots,n$, the source transmits $X^n$. We
assume $X^n$ is an i.i.d. sequence whose symbols are all independently
drawn from $\mathcal{X}$ according to $P_X$.
For description convenience, the two relays are referred to as ``relay
1" and ``relay 2".  Correspondingly, their generic received symbols
are denoted by $U_1$ and $U_2$.
Hence, 
{the channel from the source to the
      relays is characterized by the conditional PMF
      $P_{U_{1}, U_{2}\left|X\right.}$.}
We also write the joint PMF of $(U_{1},U_{2})$ in the form
of the $1\!\times\!\abs{{\mathcal{U}_1}}\abs{{\mathcal{U}_2}}$ vector
$P_{{U_1},U_2}$, whose elements are defined as:
\begin{equation}
[P_{U_{1},U_{2}}]_{j}\defn P_{U_{1},U_{2}}\left(\mathsf{u}_{1,k},
  \mathsf{u}_{2,t} \right)
\end{equation}
for  $j=\left(k-1\right)\left|\mathcal{U}_{2}\right|+t$, where $k=1,2,\ldots,
\abs{{\mathcal{U}_1}}$ and $t=1,2,\ldots, \abs{{\mathcal{U}_2}}$.

After the relays receive symbols from the source during time instants
$1,2,\ldots,n$, the relays simultaneously forward the symbols,
possibly manipulated, to the destination during the period covering
the time instants $n+1,n+2,\ldots,2n$. More precisely, relay 1 sends
the sequence $V^n_1$ while relay 2 sends the sequence $V^n_2$. We may
assume with no loss of generality that
$\mathcal{V}_{1}=\mathcal{U}_{1}$ and
$\mathcal{V}_{2}=\mathcal{U}_{2}$. Let us use $Y$ to denote the
generic symbol observed by the destination.  The multiple-access
channel from the two relays to the destination is characterized by the
conditional PMF $P_{Y\left|V_{1}V_{2}\right.}$. Like before, we may
interpret $P_{Y|V_1,V_2}$ as a
$\abs{{\mathcal{Y}}}\!\times\!\abs{{\mathcal{U}_1}}\abs{{\mathcal{U}_2}}$
matrix, whose entries are defined as:
\begin{equation}
 \left[P_{Y\left|V_{1},V_{2}\right.}\right]_{i,j} \defn P_{Y\left|V_{1},V_{2}\right.}\left(\mathsf{y}_{i}\left|
     \mathsf{v}_{1,k}, \mathsf{v}_{2,t}\right.\right),
\end{equation}
for $j=\left(k-1\right)\left|\mathcal{U}_{2}\right|+t$, where
$k=1,2,\ldots, \abs{{\mathcal{U}_1}}$ and
$t=1,2,\ldots, \abs{{\mathcal{U}_2}}$.  The PMF pair
$(P_{U_{1},U_{2}}, P_{Y\left|V_{1},V_{2}\right.})$ will be referred to
as \emph{observation channel}. The knowledge of
$(P_{U_{1},U_{2}}, P_{Y\left|V_{1},V_{2}\right.})$ is assumed to be
available to the destination for facilitating maliciousness detection.

During the period spanning the time instants $n+1,n+2,\ldots,2n$, the
destination observes the sequence $Y^{n}$. The destination needs to
determine whether the relays have manipulated their received sequences or not by processing and 
analyzing $Y^n$. In particular, we will employ the empirical
distribution $\varPi_{Y^n}$ of $Y^n$ to construct decision statistics
for detecting potential  
malicious manipulations by the relays. Note that we may again write
$\varPi_{Y^n}$ as a $1\times |\mathcal{Y}|$ vector, whose 
$j$th element is defined as
\begin{equation}\label{e:histo}
[\varPi_{Y^n}]_{j}\defn\frac{N(\mathsf{y}_{j}\left|Y^{n}\right.)}{n},
\end{equation} 
for $j=1,2,\ldots\left|\mathcal{Y}\right|$.

\subsection{Malicious Relays}
For $m=1,2$, let $U_m^n$ denote the sequence of $n$ symbols observed
by the $m$th relay during time instants $1, 2, \ldots, n$, and $V_m^n$
denote the sequence of $n$ symbols transmitted by the relay during
time instants $n+1,n+2,\ldots,2n$. Then the actions of the relays are
specified by the mapping from $(U_1^n,U_2^n,X^n)$ to $(V_1^n,V_2^n)$. We
allow this mapping to be stochastic in general, described by the
conditional PMF $P_{V_1^n,V_2^n|U_1^n,U_2^n,X^n}$, as long as it satisfies
the following Markovity constraints:
\begin{align}
P_{V_{1}^{n}|U_{1}^{n},U_{2}^{n},X^{n}}(v_{1}^{n}|u_{1}^{n},u_{2}^{n},x^{n})
&= P_{V_{1}^{n}|U_{1}^{n}}(v_{1}^{n}|u_{1}^{n})
\notag \\
 P_{V_{2}^{n}|U_{1}^{n},U_{2}^{n},X^{n}}(v_{2}^{n}|u_{1}^{n},u_{2}^{n},x^{n})
&=P_{V_{2}^{n}|U_{2}^{n}}(v_{2}^{n}|u_{2}^{n}).
\label{eq:relay}
\end{align}
Physically these constraints impose the restriction that each relay
can only formulate its attack (modification from $U^n_m$ to $V^n_m$)
based solely on the symbol sequence that it has received from the
source.

{We note that the knowledge of
$(P_{V_{1}^{n}|U_{1}^{n}}, P_{V_{2}^{n}|U_{2}^{n}})$ is usually not
available to the destination for attack detection. Thus we seek
alternative non-parametric characterizations of the relay actions that
do not require such knowledge.}
In particular, we will employ the
empirical conditional PMFs (conditional types)
$\varPi_{V_{1}^{n},V_{2}^{n}\left|U_{1}^{n},U_{2}^{n}\right.}$,
$\varPi_{V_{1}^{n}\left|U_{1}^{n}\right.}$ and
$\varPi_{V_{2}^{n}\left|U_{2}^{n}\right.}$ to characterize the actions
of relays~1 and~2. As before, for $m=1,2$, the conditional type
$\varPi_{V_{m}^{n}\left|U_{m}^{n}\right.}$ may be treated as a
$\abs{\mathcal{U}_m}\!\times\!\abs{\mathcal{U}_m}$ matrix whose
$(i,j)$th entry is defined by
\begin{equation} \label{eq:attackm}
[\varPi_{V_{m}^{n}|U_{m}^{n}}]_{i,j}
\defn
\frac{N\left(\mathsf{v}_{m,i}, \mathsf{u}_{m,j}|V_{m}^{n},U_{m}^{n}\right)}{
  N\left(\mathsf{u}_{m,j}|U_{m}^{n}\right)}.
\end{equation}
Similarly, we also treat
$\varPi_{V_{1}^{n},V_{2}^{n}\left|U_{1}^{n},U_{2}^{n}\right.}$ as a
$\abs{\mathcal{U}_1}\abs{\mathcal{U}_2}\!\times\!\abs{\mathcal{U}_1}\abs{\mathcal{U}_2}$
matrix whose $(i,j)$th entry is defined by
\begin{align} \label{eq:attackw}
\nonumber&\left[\varPi_{V_{1}^{n},V_{2}^{n}\left|U_{1}^{n},U_{2}^{n}\right.}\right]_{i,j}=\\
&\frac{N\left(\mathsf{u}_{1,k_{1}}, \mathsf{u}_{2,k_{2}},
\mathsf{v}_{1,t_{1}}, \mathsf{v}_{2,t_{2}} \left| U_{1}^{n},U_{2}^{n},V_{1}^{n},V_{2}^{n}\right.\right)}{
  N\left(\mathsf{u}_{1,k_{1}}, \mathsf{u}_{2,k_{2}}\left|U_{1}^{n},U_{2}^{n}\right.\right)},
\end{align}for $i=\left(t_{1}-1\right)\left|\mathcal{U}_{2}\right|+t_{2}$ and
$j=\left(k_{1}-1\right)\left|\mathcal{U}_{2}\right|+k_{2}$,
where $k_{1},\, t_{1}=1,2\ldots,\left|\mathcal{U}_{1}\right|$ and
$k_{2},\, t_{2}=1,2\ldots,\left|\mathcal{U}_{2}\right|$.
We will restrict ourselves to consider the family of relay actions
that satisfy the following constraint:
{\small{\begin{equation}\label{constr}
\varPi_{\!V_{1}^{n}\!,V_{2}^{n}\left|U_{1}^{n}\!,U_{2}^{n}\right.}\!\!\left(\mathsf{v}_{1}\!,\!\mathsf{v}_{2}\!\left|\mathsf{u}_{1}\!,\!\mathsf{u}_{2}\right.\!\right)\!\!\rightarrow\!\!\varPi_{\!V_{1}^{n}\left|U_{1}^{n}\right.}\!\!\left(\mathsf{v}_{1}\!\left|\mathsf{u}_{1}\right.\!\right)\!\varPi_{\!V_{2}^{n}\left|U_{2}^{n}\right.}\!\!\left(\mathsf{v}_{2}\!\left|\mathsf{u}_{2}\right.\!\right),
\end{equation}}}for all $(\mathsf{u}_1, \mathsf{v}_1) \in \mathcal{U}_1^2$ and
$(\mathsf{u}_2, \mathsf{v}_2) \in \mathcal{U}_2^2$.  This constraint
can be interpreted as the relays cannot decide their own actions by colluding
with each other, either beforehand or during the transmission process.

{We will see from Theorem 1 of Section III that for the purpose of attack
detection, the conditional types
$\varPi_{V_{1}^{n}\left|U_{1}^{n}\right.}$ and
$\varPi_{V_{2}^{n}\left|U_{2}^{n}\right.}$ are sufficient for
characterizing the maliciousness of any non-colluding relay actions that
satisfy~\eqref{constr}. In addition, it is clear that 
$\varPi_{V_{1}^{n}\left|U_{1}^{n}\right.}$ and
$\varPi_{V_{2}^{n}\left|U_{2}^{n}\right.}$ describe the maliciousness of individual
actions of relays~1 and~2, respectively. When
$\varPi_{V_{m}^{n}|U_{m}^{n}}=I$ (i.e., $V_m^n = U_m^n$), relay~$m$ is
intuitively non-malicious in the sense that it faithfully forwards its received
sequence to the destination.}
This motivates us to adopt the following more formal definition of
maliciousness of the relays based on $\varPi_{V_{1}^{n}|U_{1}^{n}}$
and $\varPi_{V_{2}^{n}|U_{2}^{n}}$:
\begin{define} \label{def:maliciousness} \textbf{(Non-malicious
    relay)} For $m=1,2$, relay~$m$ is said to be non-malicious if
  $\varPi_{V_{m}^{n}|U_{m}^{n}} \rightarrow I$ as $n$ approaches
  infinity. Otherwise, relay~$m$ is considered to be malicious. In
  addition, the relay network is considered to be safe if both relays
  are non-malicious, i.e., $\varPi_{V_{1}^{n}|U_{1}^{n}} \rightarrow I$
  and $\varPi_{V_{2}^{n}|U_{2}^{n}} \rightarrow I$.
\end{define}
{Note that the above definition of non-maliciousness of
  relay~$m$ is more relaxed than the strict requirement of
  $\varPi_{V_{m}^{n}|U_{m}^{n}} = I$, since it allows relay~$m$ to
  alter a negligible portion of its received symbols.  Although we adopt this
  relaxation for mathematical convenience, it has minimal practical
  implications, because the negligible altered symbols can
  be corrected by using channel coding/decoding in practice.}

For the family of relay actions that satisfy~\eqref{constr}, we will
show in Lemma~2 that
\begin{equation} \label{eq:conv}
P_{{U_{1},U_{2}}}(\varPi_{{V_{1}^{n}}\left|{U_{1}^{n}}\right.}^{T}\otimes\varPi_{{V_{2}^{n}}\left|{U_{2}^{n}}\right.}^{T})P_{Y\left|{V_{1},V_{2}}\right.}^{T}
\rightarrow \varPi_{Y^n}
\text{~~~as } n \rightarrow \infty.
\end{equation}
{Since the destination is capable of calculating
    $\varPi_{Y^n}$ directly from its observation $Y^n$, the above convergence 
    allows the destination to estimate
    $\varPi_{{V_{1}^{n}}\left|{U_{1}^{n}}\right.}$ and
    $\varPi_{{V_{2}^{n}}\left|{U_{2}^{n}}\right.}$ from $Y^n$ obtained
    in the physical layer. Then, the attack detection can be achieved
    by comparing the estimate of
    ($\varPi_{{V_{1}^{n}}\left|{U_{1}^{n}}\right.}$,
    $\varPi_{{V_{2}^{n}}\left|{U_{2}^{n}}\right.}$) to ($I$, $I$). The
    particular detection method and its performance analysis will be
    detailed in Appendix~B.}
  Furthermore, it is clear from the convergence characterized by~\eqref{eq:conv}
  that the observation channel $(P_{U_1,U_2}, P_{Y|V_1, V_2})$ is essential for determining the detectability of the attacks. We will investigate
  in more depth how $(P_{U_1,U_2}, P_{Y|V_1, V_2})$ affects the
  detection of such attacks in Section~III.


{Note that the restriction imposed by (\ref{constr}) excludes some possible attacks. 
Nevertheless, Proposition~1 in
  Section~III shows that the family of possible attacks that
    satisfy~(\ref{constr}) is in fact rather large. In particular, it
    allows one relay to conduct \emph{any} deterministic or random
    attack, provided that the other relay can only conduct certain
    stationary attacks, e.g., i.i.d. attacks.}
Compared with the existing schemes which restrict both relays to make
i.i.d. attacks, our results in Section~III extend the family of
attacks that can be detected at the destination based only on its
observations.


\section{Main Results} \label{se:result}

{From~\eqref{eq:conv}, we see that the observation channel
$(P_{U_1,U_2}, P_{Y|V_1, V_2})$ plays a critical role in determining
whether an attack from the family satisfying~\eqref{constr} can be detected by the
destination solely using its own observation $Y^n$. Specifically, this
detectability is determined by the number of stochastic matrix pairs
$(\Upsilon_1, \Upsilon_2)$, which are solutions to the matrix equation:
\begin{equation} \label{eq:matrix_eqn}
  P_{{U_{1},U_{2}}}P_{Y\left|{V_{1},V_{2}}\right.}^{T}=P_{{U_{1},U_{2}}}(\Upsilon_1\otimes
  \Upsilon_2)P_{Y\left|{V_{1},V_{2}}\right.}^{T},
\end{equation}where $\Upsilon_1$ and $\Upsilon_2$ are of dimensions 
$\left|\mathcal{U}_{1}\right|\times\left|\mathcal{U}_{1}\right|$ and
$\left|\mathcal{U}_{2}\right|\times\left|\mathcal{U}_{2}\right|$, respectively.
Recall from Definition~\ref{def:maliciousness} that relay~$m$ is
non-malicious if $\varPi_{V_{m}^{n}|U_{m}^{n}} \rightarrow I$ for
$m=1,2$. Thus, it is intuitive that if the observation channel
$(P_{U_1,U_2}, P_{Y|V_1, V_2})$ constitutes such a pair of distributions that make $(I,I)$ the unique
solution to~\eqref{eq:matrix_eqn}, then comparing $\varPi_{Y^n}$ with
$P_{{U_{1},U_{2}}}P_{Y\left|{V_{1},V_{2}}\right.}^{T}$ can tell us
whether malicious attacks have been carried out by the relays. On the
other hand, if \eqref{eq:matrix_eqn} has multiple solutions, then
there exist attacks that cannot be detected by simply comparing
$\varPi_{Y^n}$ with
$P_{{U_{1},U_{2}}}P_{Y\left|{V_{1},V_{2}}\right.}^{T}$. This intuitive
argument leads to the dichotomy of all observation channels
into the following two classes:}
\begin{define} \label{def:manipulable}
  {\textbf{(Non-manipulable observation
        channel)} The observation channel
      $\left(P_{{U_1,U_2}},\, P_{{Y}\left|{V_1,V_2}\right.}\right)$ is
      said to be non-manipulable if $(I,I)$ is the unique
      stochastic matrix pair that solves~\eqref{eq:matrix_eqn}.
      Otherwise, the observation channel is said to be manipulable.}

%
%
\end{define}

{With this dichotomy of observation channels, we can now present the main
results of this paper, which formally establish the detectability of
malicious attacks from the family that satisfies~\eqref{constr} at the destination:}
\begin{theorem}\textbf{(Maliciousness detectability)} \label{thm:main}
  For all malicious actions that satisfy~(\ref{constr}), the
  observation channel
  $\left(P_{{U_1,U_2}},\, P_{{Y}\left|{V_1,V_2}\right.}\right)$ is
  non-manipulable constitutes the necessary and sufficient condition for the
  existence of a decision statistic $D\left({Y}^{n}\right)$ which
  simultaneously has the following two properties: \\
  For any fixed sufficiently small $\delta >0$ and $\epsilon>0$, and for
  all sufficiently large $n$, we have
\begin{enumerate}
\item $\Pr \left\{ D\left({Y}^{n}\right) > \delta ~\Big| \sum_{m=1}^{2}
    \|\varPi_{V_{m}^{n}|U_{m}^{n}}-I \|_{2} > \delta \right\} \geq 1 -
  \epsilon$, whenever $\Pr \left\{ \sum_{m=1}^{2}
    \|\varPi_{V_{m}^{n}|U_{m}^{n}}-I \|_{2} > \delta \right\} > 0$,
\item $\Pr\left\{ D\left({Y}^{n}\right)> \varepsilon(\delta) ~\Big| \sum_{m=1}^{2}
    \|\varPi_{V_{m}^{n}|U_{m}^{n}}-I \|_{2} \leq \delta \right\} \leq
  \epsilon$, whenever $\Pr \left\{ \sum_{m=1}^{2}
    \|\varPi_{V_{m}^{n}|U_{m}^{n}}-I \|_{2} \leq \delta \right\} > 0$,
  where $\varepsilon (\delta) \rightarrow 0$ as $\delta \rightarrow 0$.
\end{enumerate}
\end{theorem}
Note that for a non-manipulable observation channel, properties 1) and
2) of Theorem 1 together imply that
$D\left({Y}^{n}\right)\rightarrow 0$ if and only if the relay network
is safe (see Definition~\ref{def:maliciousness}). This means that the
destination may detect an attack by simply comparing the decision
statistic $D(Y^n)$ to the threshold $0$. Then property~1 guarantees a
small miss detection probability when an attack has occurred, while
property~2 guarantees a small false alarm probability when no attack
has been carried out by the relays. On the other hand, no such
decision statistic exists for a manipulable observation channel, and
thus over this type of channels malicious attacks imposed by the relays cannot be accurately
detected based only on $Y^n$ at the destination.


We see that Theorem~\ref{thm:main} provides a definite answer to the
detectability of attacks in the family~\eqref{constr}. Its usefulness
and contribution depend on whether the family~\eqref{constr} contains
many common attacks beyond i.i.d attacks, whose detectability has been
previously investigated.  To this end, the following proposition shows
that~\eqref{constr} does include a very large class of attacks
including many common non-i.i.d. attacks.
\begin{proposition}
\label{memory_attack}
Let $(V_1^n, V_2^n, U_1^n,U_2^n)$ be jointly distributed
according to
\begin{align}\label{nonconspire}
  \nonumber&P_{V_1^n, V_2^n, U_1^n,U_2^n}(v_1^n, {v}_2^n, {u}_1^n, {u}_2^n) =\\
  &\hspace{0pt}P_{V_1^n|U_1^n}({v}_1^n| {u}_1^n) P_{V_2^n|U_2^n}({v}_2^n| {u}_2^n)
  \prod_{i=1}^n P_{U_1^n, U_2^n} (u_{1,i}, {u}_{2,i}).
\end{align}
Suppose that there exists a constant $c_{\mathsf{v_2}|\mathsf{u_2}}$
satisfying
$E\left\{ 1_i(\mathsf{v}_{2}|V_{2}^n) | U_{2}^n = u_{2}^n\right\} =
c_{\mathsf{v}_2|\mathsf{u}_2}$
for all $i$ such that $u_{2,i} = \mathsf{u}_{2}$. Then,
$\Pr\{ \varPi_{V_1^n|U_1^n}(\mathsf{v}_1 | \mathsf{u}_1) > 0 \}
\rightarrow 1$
and $P_{U_1,U_2}(\mathsf{u}_1, \mathsf{u}_2) > 0$ imply
\[
\varPi_{V_{1}^{n},V_{2}^{n}|U_{1}^{n},U_{2}^{n}}(\mathsf{v}_{1},\mathsf{v}_{2}|\mathsf{u}_{1},\mathsf{u}_{2})\rightarrow\varPi_{V_{1}^{n}|U_{1}^{n}}(\mathsf{v}_{1}|\mathsf{u}_{1})\varPi_{V_{2}^{n}|U_{2}^{n}}(\mathsf{v}_{2}|\mathsf{u}_{2}).
\]
\end{proposition}
Note that Proposition~\ref{memory_attack} only restricts one relay's
(relay~2) malicious action, and the other relay's (relay 1) attack can
be arbitrary without any restriction. The condition
$E\left\{ 1_i(\mathsf{v}_{2}|V_{2}^n) | U_{2}^n = u_{2}^n\right\} =
c_{\mathsf{v}_2|\mathsf{u}_2}$
for relay~2 is a first-order stationarity requirement on the attack
mapping from $U_2^n$ to $V_2^n$. It is easy to check that this
condition is satisfied by i.i.d. attacks as well as any attack that
can be modeled by a Markov chain with a unique stationary
distribution. 



\begin{figure}
\centering
\includegraphics[width=0.45\textwidth]{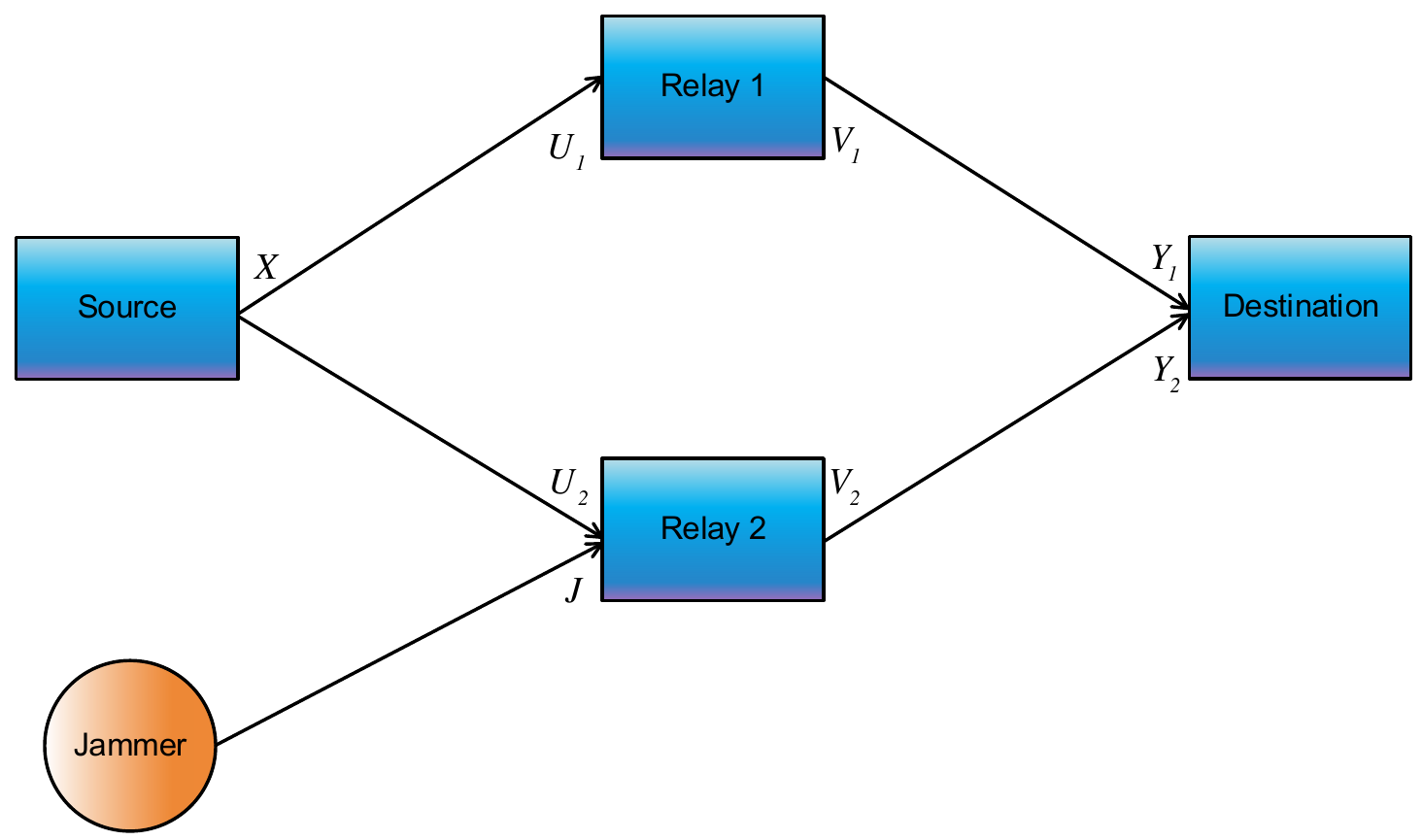}
\caption{A relay network with a jammer, which transmits $J^{n}$ obeying a stationary Markov stochastic process.}
\label{fig:memory}
\end{figure}
The example illustrated by Fig.~\ref{fig:memory} describes an attack
scenario that can be modeled by a stationary Markov chain, and hence
by Proposition~\ref{memory_attack} satisfying~\eqref{constr}. More
specifically, in Fig.~\ref{fig:memory} a jammer inserts malicious
sequence $J^{n}$ into the data flowing to relay~2. 
  {We assume that the symbols of $J^n$, $V^n_2$ and
    $U^n_2$ take values from the same finite field
    $\mathbb{F}_{q}$. The $i$th output symbol of relay~2 is given by
    $V_{2,i}=\left(J_i+U_{2,i}\right)_{q}$, where
    $\left(\cdot\right)_{q}$ is the modulo value of its argument with
    base~$q$, and $q$ is a prime number.}
This assumption approximately models the case that the symbols of
$J^n$ and $U^n_2$ superimpose before they arrive at relay~2, which
simply forwards its received symbols to the destination.
We assume that the jammer sequence $J^{n}$ is a stationary Markov
sequence satisfying $P_{J_{i}}=P_{J_{i}}P_{J_{i+1}|J_{i}}$,
$i=1,\ldots n$, where $P_{J_{i}}$ is the distribution of $J_{i}$, and
$P_{J_{i+1}|J_{i}}$ specifies the transfer probability from ${J_{i}}$
to ${J_{i+1}}$. In addition, $J^n$ is also independent of all other
random variables. Note that the action of relay~1 specified by the
mapping from $U^n_1$ to $V^n_1$ is arbitrary.
It is easy to check that for all $i$ such that
$u_{2,i} = \mathsf{u}_{2}$,
$E\left\{1_{i}(\mathsf{v}_{2}|V_{2}^{n})|U_{2}^{n}=u_{2}^{n}\right\}
=P_{J_{i}}\left(\left(\mathsf{v}_{2}-\mathsf{u}_{2}\right)_{q}\right)$.
Therefore, Proposition~\ref{memory_attack} applies, showing that this
attack scenario is within the family~\eqref{constr}.  Theorem
\ref{thm:main} is thus applicable to this attack scenario as well.

In summary, Theorem~\ref{thm:main} and Proposition~\ref{memory_attack}
together show that non-manipulability of the observation channel
$(P_{U_1,U_2}, P_{Y|V_1,V_2})$ is necessary and sufficient for the
detectability of a very large family of attacks by solely using the
sequence 
received at the destination. This family of attacks is specified
by~\eqref{constr} and includes attacks in which the action of one
relay can be modeled by a first-order stationary mapping, while the
action of the other relay can be completely arbitrary.
This contribution distinguishes our work from the
existing literature that either focuses on i.i.d. attacks conducted by
both relays, or imposes the requirement of utilizing 
reliable observations. 

We note that while the family~\eqref{constr} contains a very large
class of attacks that can be detected at the destination by
Theorem~\ref{thm:main}, Proposition~\ref{permutation} below shows that
there are attacks that cannot be detected based on observing only
$Y^n$.
\begin{proposition}
\label{permutation}
Let $(V_1^n, V_2^n, U_1^n,U_2^n)$ be distributed according
to~(\ref{nonconspire}).  Consider the attack in which the two relays
permutate their received symbols according to the same pre-determined
order. This means that $V_{1}^{n}$ and $V_{2}^{n}$ are respective
rearrangements of $U_{1}^{n}$ and $U_{2}^{n}$, and for any
$k,t=1,\ldots n$, if $V_{1,k}=U_{1,t}$ and $V_{1,t}=U_{1,k}$, we must
have $V_{2,k}=U_{2,t}$ and $V_{2,t}=U_{2,k}$. For this attack, there
does not exist any decision statistic $D\left({Y}^{n}\right)$ that 
satisfies the two properties given in Theorem~\ref{thm:main}.
\end{proposition}
Note that the permutation attack described in
Proposition~\ref{permutation} does not belong to the
family~\eqref{constr}. Intuitively, the two relays collude in
agreeing to permute their respective received symbols in the same
order. We see that the lack of a clean reference at the destination
prevents it from detecting this kind of attacks in which the two
relays collude.


  {To summarize, we have briefly discussed some practical
    implications of the theoretical detectability results provided by
    Theorem~1, Proposition~1 and Proposition~2.  These results show that it is
    possible to perform Byzantine attack detection when no clean
    reference is available at the destination for a large family of
    non-colluding attacks.  The lack of a clean reference is rather
    commonplace in many peer-to-peer or ad hoc networks, where we
    may not be able to guarantee the trustworthiness of all relay
    nodes. The Byzantine attack detectability results in Theorem~1 and
    Proposition~1 suggest us to form the network in a more judicious
    manner so that it is less likely to select relay nodes that may
    collude in attack. By doing so, malicious attacks by the relay nodes
    can be detected. Moreover, since the proposed Byzantine attack detection
    techniques do not require any cryptographic aid, the usual
    overheads incurred in the implementation of cryptographic systems,
    such as key management and distribution, can be avoided.}

\section{Checking the non-manipulability condition}
As shown in Theorem 1, the detectability of attacks in the
family~\eqref{constr} is determined by whether the condition of
non-manipulability holds for the observation channel
$(P_{U_1,U_2}, P_{Y|V_1,V_2})$ or not. Recall that
$(P_{U_1,U_2}, P_{Y|V_1,V_2})$ is non-manipulable if $(I,I)$ is the
unique stochastic solution to~\eqref{eq:matrix_eqn}. This condition is
sometimes difficult to check.  In this section, we discuss a
numerically efficient method to check the non-manipulability
condition.
{The main idea is to transform the task of checking
    the condition of non-manipulability into an optimization problem
    that can be solved by standard methods. To this end, we first
    rewrite the solution of~\eqref{eq:matrix_eqn} as that of an
    optimization problem (see \eqref{non_convex} below). The
    non-manipulability condition is equivalent to the solution to the
    optimization problem taking an extremal value in the possible
    range. This optimization problem is non-convex as its feasible set
    is non-convex.  In order to find numerically efficient methods to
    solve this optimization problem, we further relax the feasible set
    to a convex set, which results a convex optimization problem (see
    \eqref{convex} below) that can be solved efficiently.  }

More specifically, we first consider the optimization below:
\begin{subequations}\label{non_convex}
\begin{align}
\min_{\begin{subarray}{c}
\Upsilon_1,\Upsilon_2\end{subarray}} & \sum_{k=1}^{\left|\mathcal{U}_{1}\right|}\left[\Upsilon_{1}\right]_{k,k}+\sum_{k=2}^{\left|\mathcal{U}_{2}\right|}\left[\Upsilon_{2}\right]_{k,k}\\
s.t. \: & P_{{U_{1},U_{2}}}(\Upsilon_{1}^{T}\otimes \Upsilon_{2}^{T})P_{Y\left|{V_{1},V_{2}}\right.}^{T}=P_{{U_{1},U_{2}}}P_{Y\left|{V_{1},V_{2}}\right.}^{T},\\
 & 0\leq\left[\Upsilon_{1}\right]_{i,j}\leq1,\quad1\leq i\leq\left|\mathcal{U}_{1}\right|,1\leq j\leq\left|\mathcal{U}_{1}\right|,\\
 & 0\leq\left[\Upsilon_{2}\right]_{i,j}\leq1,\quad1\leq i\leq\left|\mathcal{U}_{2}\right|,1\leq j\leq\left|\mathcal{U}_{2}\right|,\\
 &
   \sum_{i=1}^{\left|\mathcal{U}_{1}\right|}\left[\Upsilon_{1}\right]_{i,j}=1,\quad j=1,2,\ldots,\left|\mathcal{U}_{1}\right|,\\
& \sum_{i=1}^{\left|\mathcal{U}_{2}\right|}\left[\Upsilon_{2}\right]_{i,j}=1,\quad j=1,2,\ldots,\left|\mathcal{U}_{2}\right|.
\end{align}
\end{subequations}
It is clear that the minimum value of the problem~(\ref{non_convex})
lies inside the interval
$\left[0,\left|\mathcal{U}_{1}\right|+\left|\mathcal{U}_{2}\right|\right]$.
Moreover, under the
constraints~(\ref{non_convex}b)--(\ref{non_convex}f), the value of the
objective function in~(\ref{non_convex}) can be equal to
$\left|\mathcal{U}_{1}\right|+\left|\mathcal{U}_{2}\right|$ if and
only if $\Upsilon_1=I$ and $\Upsilon_2=I$. For any other choice of
$(\Upsilon_1,\Upsilon_2)$, the value of the objective function
in~(\ref{non_convex}) is strictly smaller than
$\left|\mathcal{U}_{1}\right|+\left|\mathcal{U}_{2}\right|$. From this
observation, we can conclude that $(\Upsilon_1,\Upsilon_2)=(I,I)$ is
the unique solution to~\eqref{eq:matrix_eqn}, and hence
$\left(P_{{U_1,U_2}},\, P_{{Y}\left|{V_1,V_2}\right.}\right)$ is
non-manipulable, if and only if the minimum value of the
problem~(\ref{non_convex}) is
$\left|\mathcal{U}_{1}\right|+\left|\mathcal{U}_{2}\right|$. Thus we
may determine whether the observation channel is non-manipulable by
solving the optimization problem~(\ref{non_convex}) and checking
whether the minimum value attained is
$\left|\mathcal{U}_{1}\right|+\left|\mathcal{U}_{2}\right|$.  Because
the constraint (\ref{non_convex}b) is not convex, the optimization
problem~(\ref{non_convex}) is non-convex. Hence, it is challenging to
solve the problem in a computationally efficient manner.

To check the non-manipulability condition efficiently, we relax the
feasible set specified by~(\ref{non_convex}b)--(\ref{non_convex}f)
into a convex set in order to obtain
convexity. Note that the non-convexity of
(\ref{non_convex}) is caused by the Kronecker product appearing
in~(\ref{non_convex}b). In order to get a convex feasible set, we
replace $\Upsilon_{1}\otimes \Upsilon_{2}$ with a
$\left|\mathcal{U}_{1}\right|\left|\mathcal{U}_{2}\right|\times\left|\mathcal{U}_{1}\right|\left|\mathcal{U}_{2}\right|$
matrix $\Upsilon$ in (\ref{non_convex}b).
 Then from the constraint that
$\Upsilon_{1}$ and $\Upsilon_{2}$ are
 stochastic matrices, $W$ must satisfy the following linear
 constraints:
{\small{ \begin{align}
 \label{L1}
&\hspace{0pt} \sum_{i=1}^{\left|\mathcal{U}_{1}\right|\left|\mathcal{U}_{2}\right|}\left[\Upsilon\right]_{i,j}=1,\quad j=1,2,\ldots,\left|\mathcal{U}_{1}\right|\left|\mathcal{U}_{2}\right|,\\
\label{L2}
&\hspace{0pt} \sum_{i=1}^{\left|\mathcal{U}_{1}\right|}\left[\Upsilon_{1}\right]_{i,j}=1,\quad j=1,2,\ldots,\left|\mathcal{U}_{1}\right|,\\
\label{L3}
\nonumber&\sum_{i=\left(t-1\right)\left|\mathcal{U}_{2}\right|+1}^{t\left|\mathcal{U}_{2}\right|}\left[\Upsilon\right]_{i,\left(k-1\right)\left|\mathcal{U}_{2}\right|+1}=\sum_{i=\left(t-1\right)\left|\mathcal{U}_{2}\right|+1}^{t\left|\mathcal{U}_{2}\right|}\left[\Upsilon\right]_{i,\left(k-1\right)\left|\mathcal{U}_{2}\right|+2}\\
&=\cdots
=\!\!\sum_{i=\left(t-1\right)\left|\mathcal{U}_{2}\right|+1}^{t\left|\mathcal{U}_{2}\right|}\!\!\left[\Upsilon\right]_{i,k\left|\mathcal{U}_{2}\right|}=\left[\Upsilon_{1}\right]_{t,k},\; t,k=1,2\ldots\left|\mathcal{U}_{1}\right|,\\
\label{L4}
\nonumber&\sum_{i=0}^{\left|\mathcal{U}_{1}\right|-1}\left[\Upsilon\right]_{i\left|\mathcal{U}_{2}\right|+t,k}=\sum_{i=0}^{\left|\mathcal{U}_{1}\right|-1}\left[\Upsilon\right]_{i\left|\mathcal{U}_{2}\right|+t,k+\left|\mathcal{U}_{2}\right|}\\
\nonumber&=\cdots=\sum_{i=0}^{\left|\mathcal{U}_{1}\right|-1}\left[\Upsilon\right]_{i\left|\mathcal{U}_{2}\right|+t,k+\left(\left|\mathcal{U}_{1}\right|-1\right)\left|\mathcal{U}_{2}\right|}=\left[\Upsilon_{2}\right]_{t,k},\;\\ &\hspace{150pt}t,k=1,2,\ldots\left|\mathcal{U}_{2}\right|.
 \end{align}}}
Adding these four linear constraints back,
the feasible set specified by (\ref{non_convex}b)-(\ref{non_convex}f) can be relaxed, and correspondingly
we obtain the following linear optimization problem which is a relaxation
of~\eqref{non_convex}:
\begin{subequations}\label{convex}
\begin{align}
\min_{\begin{subarray}{c}
\Upsilon_1,\Upsilon_2\end{subarray}} & \sum_{k=1}^{\left|\mathcal{U}_{1}\right|}\left[\Upsilon_{1}\right]_{k,k}+\sum_{k=2}^{\left|\mathcal{U}_{2}\right|}\left[\Upsilon_{2}\right]_{k,k}\\
s.t. \: & P_{{U_{1},U_{2}}}\Upsilon^{T}P_{Y\left|{V_{1},V_{2}}\right.}^{T}=P_{{U_{1},U_{2}}}P_{Y\left|{V_{1},V_{2}}\right.}^{T},\\
 & 0\leq\left[\Upsilon_{1}\right]_{i,j}\leq1,\quad1\leq i\leq\left|\mathcal{U}_{1}\right|,1\leq j\leq\left|\mathcal{U}_{1}\right|,\\
 & 0\leq\left[\Upsilon_{2}\right]_{i,j}\leq1,\quad1\leq i\leq\left|\mathcal{U}_{2}\right|,1\leq j\leq\left|\mathcal{U}_{2}\right|,\\
 & (\ref{L1}), (\ref{L2}), (\ref{L3}), \text{ and } (\ref{L4}).
\end{align}
\end{subequations}
We note that this linear program can be solved efficiently using
standard linear programming techniques.  Comparing the optimization
problems (\ref{non_convex}) and (\ref{convex}), it is clear that both
have the same objective function. Furthermore, since the feasible set
of (\ref{non_convex}) is a subset of that of (\ref{convex}), the
minimum value of (\ref{convex}) must be no greater than the minimum
value of (\ref{non_convex}). As a result, if the minimum value of
(\ref{convex}) is
$\left|\mathcal{U}_{1}\right|+\left|\mathcal{U}_{2}\right|$, then the
minimum value of (\ref{non_convex}) must also be
$\left|\mathcal{U}_{1}\right|+\left|\mathcal{U}_{2}\right|$, and hence
the observation channel
$\left(P_{{U_1,U_2}},\, P_{{Y}\left|{V_1,V_2}\right.}\right)$ is
non-manipulable.
Although the converse is not true for the linear
program~\eqref{convex}, we may still use it to check for
non-manipulability of the observation channel first before opting to
solve the non-convex problem~\eqref{non_convex}.

\section{Numerical Examples} \label{se:examples}

We give three numerical examples in this section to illustrate the
detectability results of Theorem~\ref{thm:main}, Propositions~1
and~2. In all the examples, we assume that the source alphabet is
binary (i.e., $\abs{\mathcal{X}}=2$) and all the other alphabets are
ternary (i.e.,
$\abs{\mathcal{U}_1} = \abs{\mathcal{U}_2} = \abs{\mathcal{V}_1} =
\abs{\mathcal{V}_2} = 3$).
For simplicity, we also assume $P_{Y\left|{V_{1},V_{2}}\right.}=I$ and
$P_{{U_{1},U_{2}\left|X\right.}}=P_{{U_{1}\left|X\right.}}\otimes
P_{{U_{2}\left|X\right.}}$.
In addition, we note that the decision statistic used in the proof of
Theorem~\ref{thm:main} (see Appendix~\ref{se:suffproof}) is too
complicated for practical implementation. Therefore, we employ the
simple heuristic decision statistic
\begin{equation}
  D\left({Y}^{n}\right)=\left\|\varPi_{Y^{n}}-P_{{U_{1},U_{2}}}P_{Y\left|{V_{1},V_{2}}\right.}^{T}\right\|_2
\label{eq:pracest}
\end{equation}
for Byzantine attack detection in the examples.

\subsection{A Non-manipulable Observation Channel Example for
  Demonstrating the Sufficiency in Theorem 1}
We first choose the source symbol distribution
$P_{X}=\left[\begin{array}{cc} .4999 & .5001\end{array}\right]$. Then,
we consider a relay network in which the channels from the
source to the two relays are specified by the PMF matrices
$P_{U_{1}|X}=P_{U_{2}|X}=\left[\begin{array}{cc}
                                 .9 & 0\\
                                 .1 & .1 \\
                                 0 & .9
\end{array}\right]$,
while the channels from the two relays to the destination are assumed
to be perfect, i.e., $P_{Y\left|{V_{1},V_{2}}\right.}=I$. By solving
the linear program~\eqref{convex}, it is easy to check that the
observation channel $(P_{U_{1},U_{2}},I)$ in this example is
non-manipulable.

\begin{figure}
\centering
\includegraphics[width=0.48\textwidth]{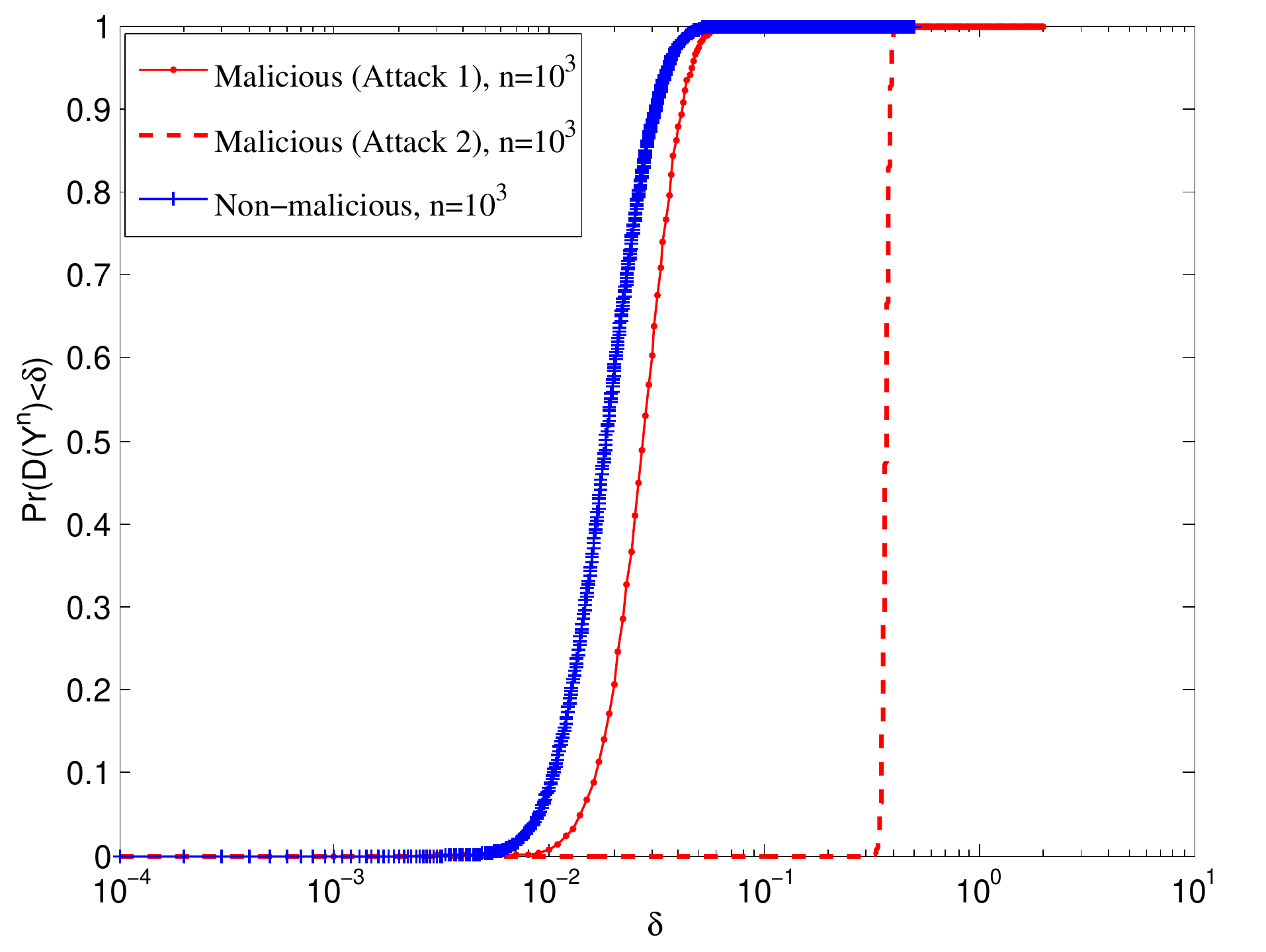}
\caption{Empirical CDFs of $D\left({Y}^{n}\right)$, i.e.,
  $\Pr (D\left({Y}^{n}\right) \leq \delta)$, in the non-manipulable
  observation channel
  example considered in Section V-A. For $n = 10^3$, three different
  cases are compared, namely, the ``non-malicious'' case as well as
  the malicious cases of Attack 1 and Attack 2.}
\label{fig:1e3}
\end{figure}
\begin{figure}
\centering
\includegraphics[width=0.48\textwidth]{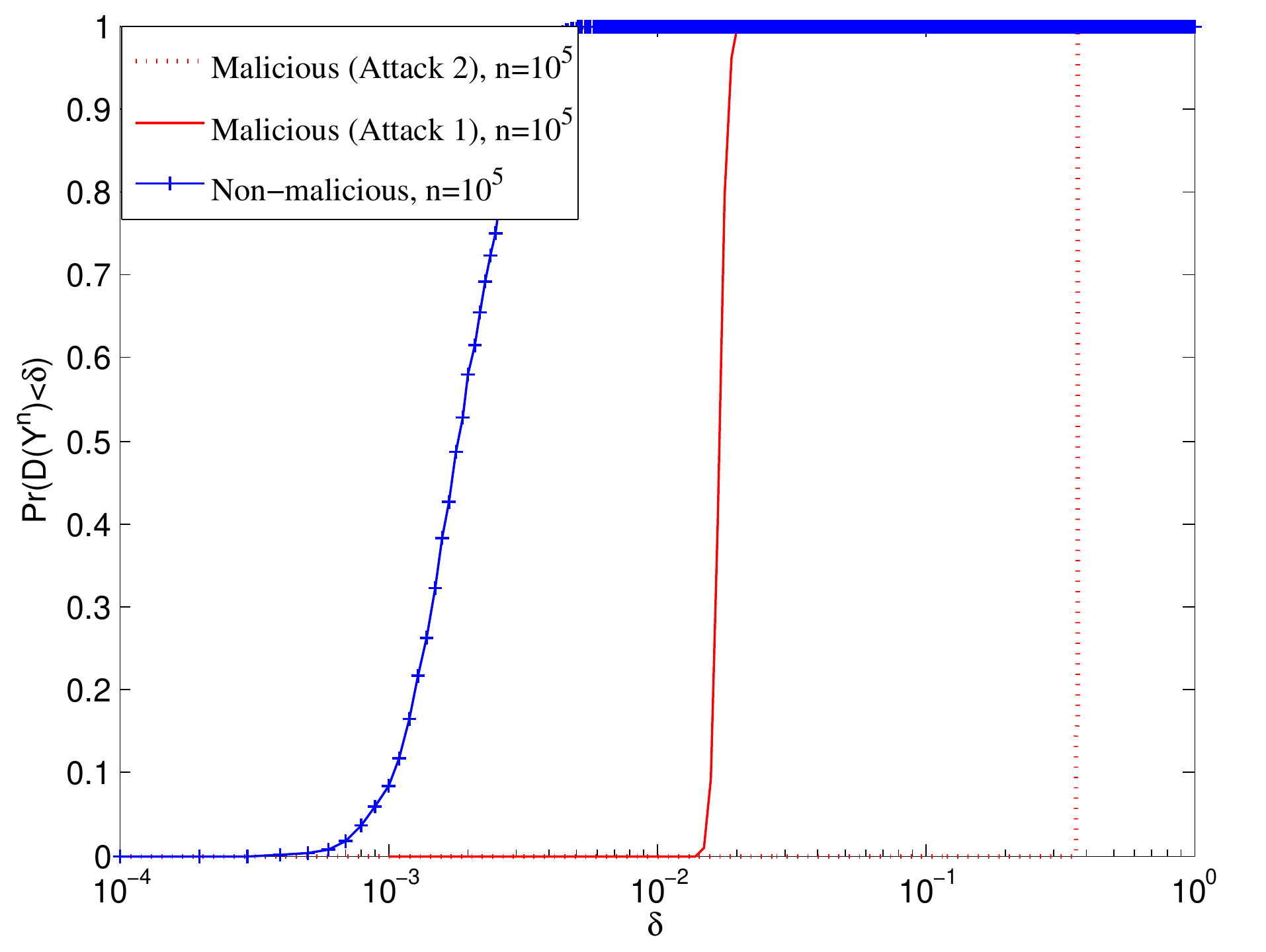}
\caption{Empirical CDFs of $D\left({Y}^{n}\right)$, i.e.,
  $\Pr (D\left({Y}^{n}\right) \leq\delta)$, in the non-manipulable
  observation channel example
  considered in Section V-A. For $n = 10^5$, three different cases
  are compared, namely, the ``non-malicious'' case as well as the
  malicious cases of Attack 1 and Attack 2.}
\label{fig:1e5}
\end{figure}

Two different attacks are considered.  In the first attack, referred
to as Attack 1, relay~2 performs an i.i.d. attack by randomly and
independently switching its received symbols according to the
conditional distribution specified by
$P_{V_{2}\left|U_{2}\right.}=\left[\begin{array}{ccc}
                                     .995 & .0025 & .0025\\
                                     .0025 & .995 & .0025\\
                                     .0025 & .0025 & .995
\end{array}\right]$, while relay~1 conducts a non-i.i.d. attack by
randomly and independently switching its $i$th input symbol $U_{1,i}$
according to the conditional distribution
$P_{V_{1}\left|U_{1}\right.}=\left[\begin{array}{ccc}
                                     .995 & .0025 & .0025\\
                                     .0025 & .995 & .0025\\
                                     .0025 & .0025 & .995
\end{array}\right]$
when $i$ is odd, and according to
$\tilde P_{V_{1}\left|U_{1}\right.}=\left[\begin{array}{ccc}
                                            .95 & .025 & 0\\
                                            .05 & .95 & .05\\
                                            0 & .025 & .95
\end{array}\right]$
when $i$ is even. Since relay~2 performs an i.i.d. attack, Attack~1 is
in the family~\eqref{constr} according to Proposition~1.
By Theorem 1, Attack 1 can be detected in the non-manipulable
observation channel $(P_{U_1,U_2}, I)$.

In the second attack considered, referred to as Attack 2, relay~1
switches its input in the same manner as it does in
Attack~1. Meanwhile,  relay~2 conducts the stationary Markov attack as
illustrated in Fig. 2, with
$P_{J_{1}}=\left[\begin{array}{ccc} \frac{6}{25}, & \frac{2}{5}, &
    \frac{9}{25}\end{array}\right]$ and
$P_{J_{i+1}|J_{i}}=\left[\begin{array}{ccc}
                           \frac{1}{3} & \frac{1}{3} & \frac{1}{3}\\
                           \frac{1}{4} & \frac{1}{2} & \frac{1}{4}\\
                           \frac{1}{6} & \frac{1}{3} & \frac{1}{2}
\end{array}\right]$, $i=1,2\ldots,n$.
As explained before (see the description of Fig.~2 in Section~III),
Proposition~1 gives that Attack~2 belongs to the
family~\eqref{constr}.
Thus by Theorem~1, Attack~2 can also be detected in the
non-manipulable observation channel.

We conduct computer simulations to illustrate the detectability of both
attacks using the heuristic decision statistic $D\left({Y}^{n}\right)$
in~\eqref {eq:pracest}.  The empirical cumulative distribution
functions (CDFs) of $D\left({Y}^{n}\right)$ obtained from the
simulation are plotted in Figs.~\ref{fig:1e3} and~\ref{fig:1e5} for
the cases of $n=10^3$ and $n=10^5$, respectively. From
Fig.~\ref{fig:1e5}, it is observed that there are clear separations between the
empirical CDFs of $D\left({Y}^{n}\right)$ for the non-malicious case
and the malicious cases of Attack 1 and Attack 2 when $n=10^5$.
For instance, as seen in Fig. 4, using the decision threshold
$\delta=0.01$ for $D\left({Y}^{n}\right)$, one may differentiate
between the non-malicious (both relays faithfully forward their
respective received sequences to the destination) and malicious cases
with negligible miss detection and false alarm probabilities. This
observation verifies sufficiency of non-manipulability for the
detectability of Byzantine attacks in the family~\eqref{constr}
promised by Theorem 1.

\subsection{A Manipulable Observation Channel Example for
  Demonstrating the Necessity in Theorem~\ref{thm:main}}
\begin{figure}
\centering
\includegraphics[width=0.48\textwidth]{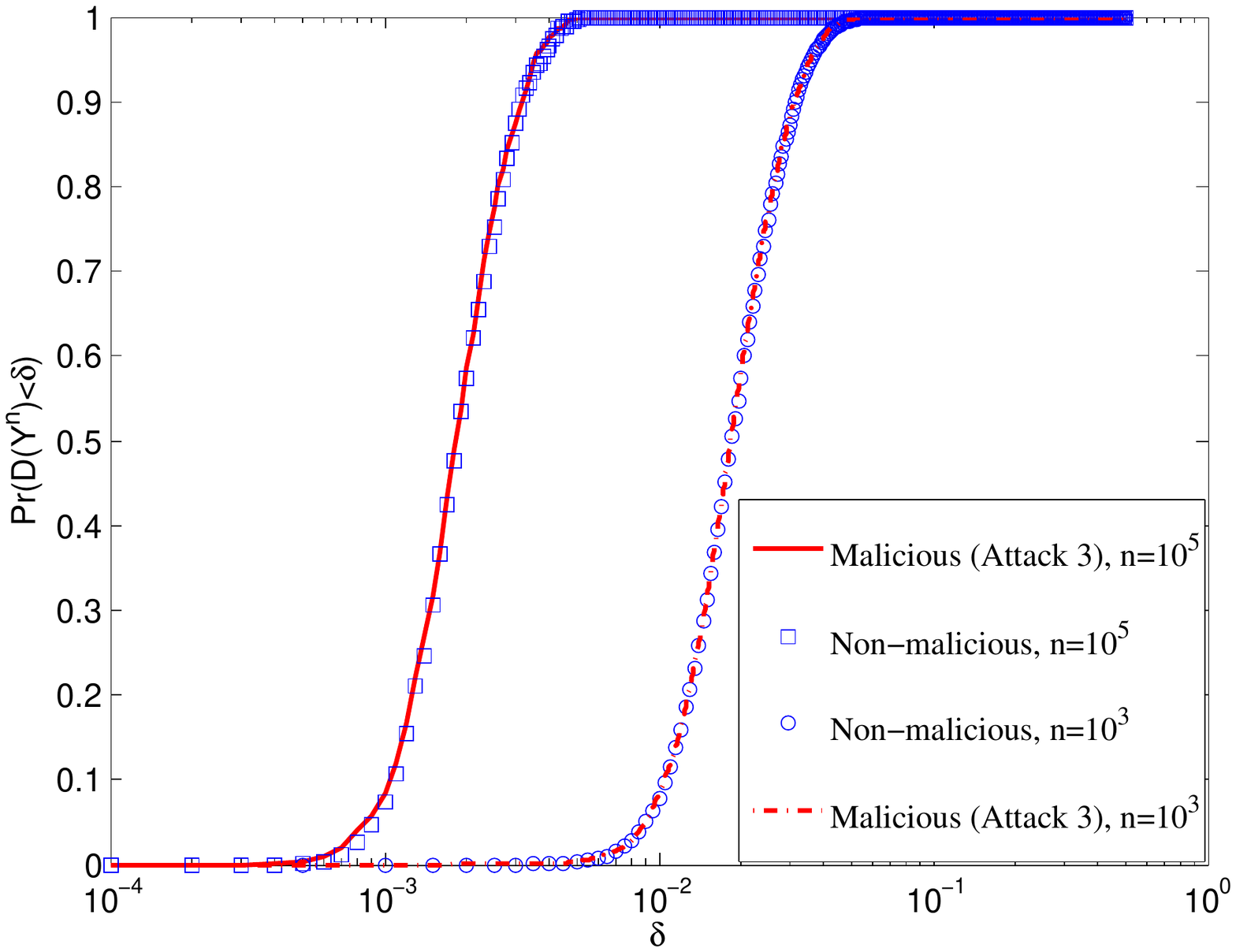}
\caption{Empirical CDFs of $D(Y^n)$, i.e., $\Pr (D^n \leq\delta)$, in
  the manipulable observation channel example considered in Section
  V-B.}
\label{fig:manipulable}
\end{figure}

We again consider the same relay network used in the previous
example, except that the source symbol distribution is now
$P_{X}=\left[\begin{array}{cc}
.5 & .5\end{array}\right]$ and hence giving
$P_{U_{1},U_{2}}=\left[\begin{array}{ccc}
    .125 & .125 & 0\\
    .125 & .25 & .125\\
    0 & .125 & .125
\end{array}\right]$.
It is straightforward to check that the choice of
$\Upsilon_{1}=\Upsilon_{2}=\left[\begin{array}{ccc}
    -1 & 0 & 1\\
    0 & 0 & 0\\
    1 & 0 & -1
\end{array}\right]$ makes this observation channel
$(P_{U_{1},U_{2}},I)$ manipulable. Therefore, according to
Theorem~\ref{thm:main}, we know that the attacks in the
family~\eqref{constr} are not detectable in this network. In order to
verify this, a simulation is carried out for an i.i.d. attack,
referred to as Attack 3, where each relay randomly and independently
switches its input symbols according to the conditional distributions
$P_{V_{1}\left|U_{1}\right.}=P_{V_{2}\left|U_{2}\right.}=\left[\begin{array}{ccc}
                                                                 0 & 0 & 1\\
                                                                 0 & 1 & 0\\
                                                                 1 & 0
                                                                       &
                                                                         0
\end{array}\right]$. Attack 3
corresponds to the scenario where each relay flips the first and third
symbols in their alphabet. The empirical CDFs of
$D\left({Y}^{n}\right)$ for the ``non-malicious" case and this
malicious case obtained from the simulation are plotted in
Fig.~\ref{fig:manipulable}. It is clearly seen from Fig. 5 that
the CDFs are indistinguishable, regardless of the value of $n$. As a
result, from $D\left({Y}^{n}\right)$, we cannot differentiate between
the non-malicious and malicious cases.  This observation indicates that the
two properties of Theorem~1 cannot be simultaneously satisfied in this
manipulable observation channel.

\subsection{An Example for Demonstrating Proposition 2}
\begin{figure}
\centering
\includegraphics[width=0.48\textwidth]{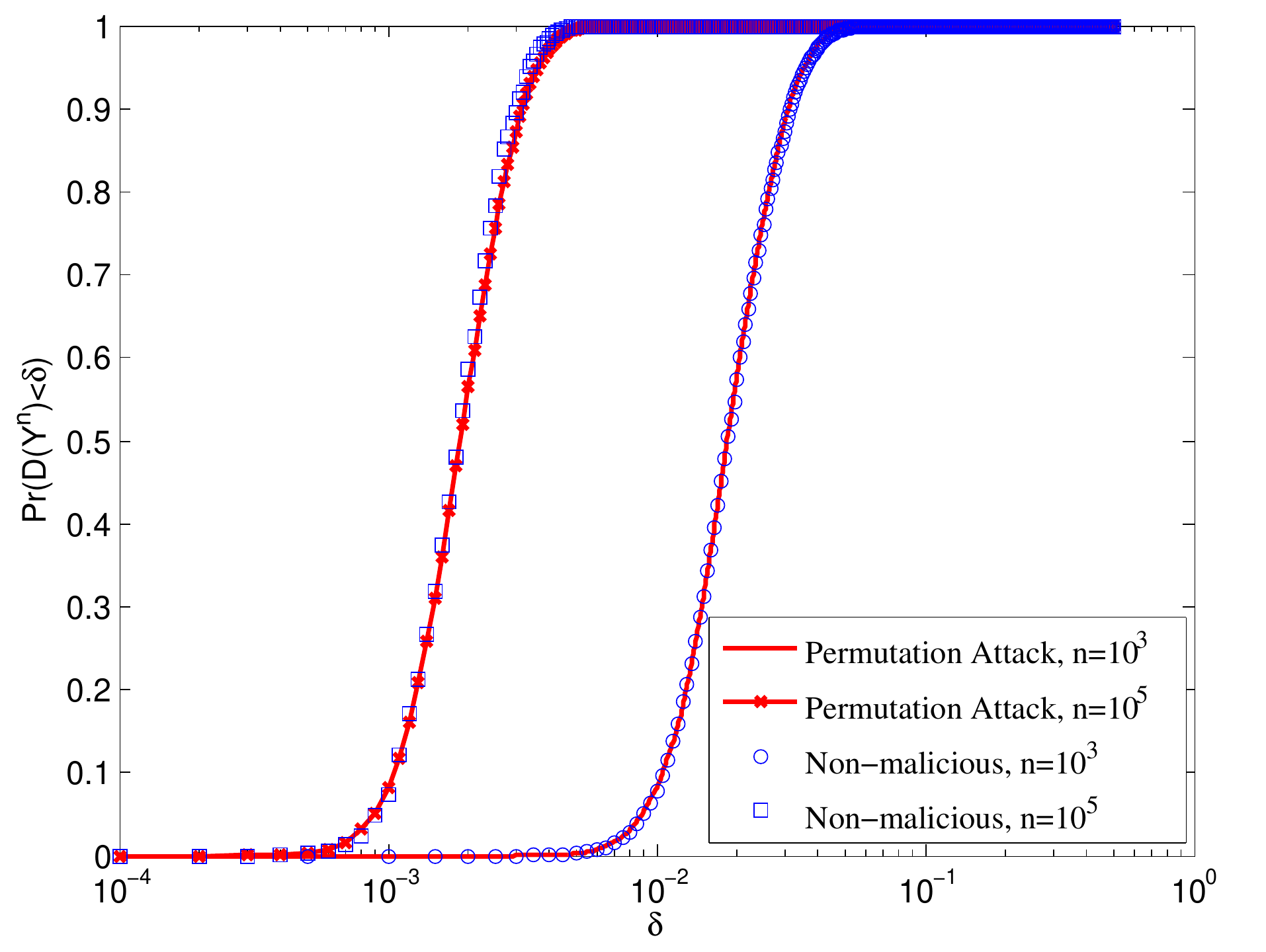}
\caption{Empirical CDFs of $D\left({Y}^{n}\right)$, i.e., $\Pr
  (D\left({Y}^{n}\right)<\delta)$, for the non-malicious case and the
  malicious case of  the permutation attack in the non-manipulable
  observation channel of Section V-A.}
\label{fig:p1}
\end{figure}

\begin{figure}
\centering
\includegraphics[width=0.48\textwidth]{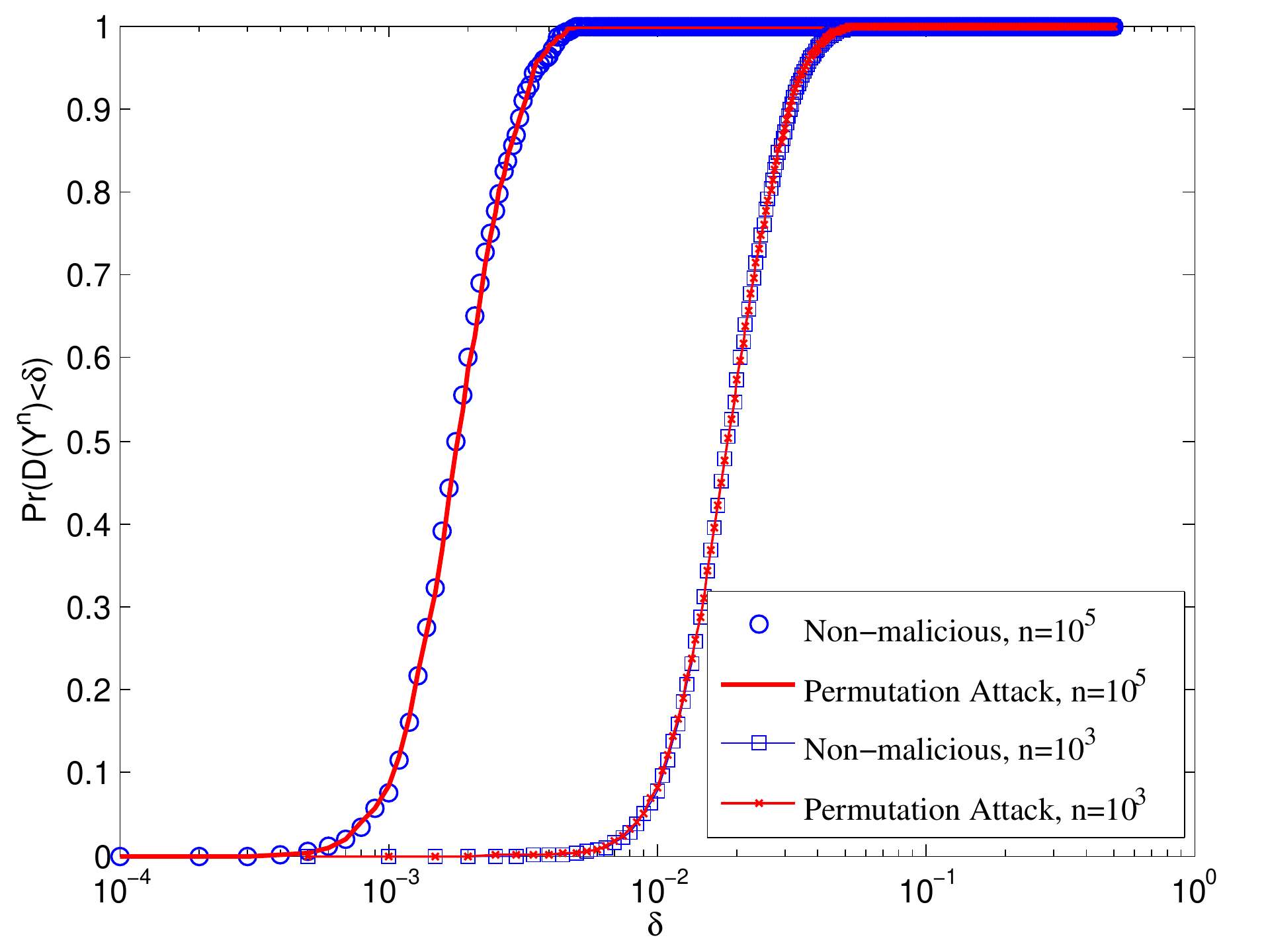}
\caption{Empirical CDFs of $D\left({Y}^{n}\right)$, i.e., $\Pr
  (D\left({Y}^{n}\right) \leq\delta)$ for the non-malicious case and
  the malicious case of the permutation attack in the manipulable
  observation channel of Section V-B.}
\label{fig:p2}
\end{figure}

In order to verify the assertion of Proposition 2, we consider the
permutation attack in which each relay permutes the symbols in its
input sequences according to
$v_{m,i}=u_{m,\left(i+\frac{n}{2}\right)_{n}}$, where $m=1,2$, and
$\left(\cdot\right)_{n}$ is the modulo value of its argument with base
$n$.  Simulations are conducted for the permutation attack in both the
non-manipulable observation channel and the manipulable observation
channel considered previously in Sections V-A and V-B,
respectively. The empirical CDFs of $D\left({Y}^{n}\right)$ for the
non-malicious case and the malicious case of the permutation attack
obtained from the simulations are plotted in Fig.~\ref{fig:p1} and
Fig.~\ref{fig:p2} for the non-manipulable and manipulable observation
channels, respectively. We can clearly see from Figs.~6 and~7 that the
CDFs for the malicious and non-malicious cases are indistinguishable,
regardless of the value of $n$ and whether the observation channel is
manipulable or not.  In other words, the permutation attack is not
detectable. This is consistent with the assertion of Proposition 2.

\section{Conclusions} \label{se:conc}
We have considered the detectability of Byzantine attacks conducted by
two non-colluding relay nodes in a relay network, where there are two
independent paths, each via one relay node, from the source to the
destination. No clean reference is assumed available to the
destination to detect whether Byzantine attacks have been carried out
by the relay nodes. The destination is allowed to employ only the symbol
sequence that it has received from the relays to perform attack
detection.  We have identified a family of attacks that can be
detected with asymptotically small miss detection and false alarm
probabilities if and only if the channels that make up the relay
network satisfies a non-manipulability condition. This family of
attacks physically correspond to ones in which the two relays do not
collude in attack. In addition, we have
also shown that there are attacks, which do not belong to this family,
that are not detectable by the destination due to the lack of a clean
reference. These results provide the practical insight that we should
choose relay nodes judiciously in order to reduce the possibility that
they may collude in attack. 
\appendices{}
\section{Proof of Proposition \ref{memory_attack} and Proposition \ref{permutation}}
 \subsection{Proof of Proposition \ref{memory_attack}}
In order to prove Proposition \ref{memory_attack}, we first give the following lemma.
\begin{lemma}\label{lem5}
  Let $\tilde V^n$, $V^n$, $U^n$ and $W^n$ be jointly distributed
  random sequences satisfying:
\begin{enumerate}
\item $\Pr\{ \varPi_{U^n, V^n}(\mathsf{u}, \mathsf{v}) > 0 \} \rightarrow 1$,
\item $(\tilde V^n,V^n)$ and $W^n$ are conditionally independent given
  $U^n$, and
\item there exists a constant $c_{\mathsf{w}|\mathsf{u}}$ such that
  $E\left\{ 1_i(\mathsf{w}|W^n) | U^n = u^n\right\}  = c_{\mathsf{w}|\mathsf{u}}$
  for all $u_i = \mathsf{u}$.\emph{}
\end{enumerate}
Then, we have
$\varPi_{W^n|U^n,
  V^n}(\mathsf{w}| \mathsf{u}, \mathsf{v}) \rightarrow \varPi_{W^n|
  U^n}(\mathsf{w}|\mathsf{u})$ for all $\mathsf{w}$.
Furthermore, if 
\begin{enumerate}
\item[4)] $\Pr\{\varPi_{W^n|U^n}(\mathsf{w} | \mathsf{u}) > 0\}
\rightarrow 1$
\end{enumerate}is also satisfied, then $\varPi_{\tilde V^n|U^n,
  V^n,W^n}(\mathsf{\tilde v}| \mathsf{u}, \mathsf{v}, \mathsf{w})
\rightarrow \varPi_{\tilde V^n|U^n, V^n}( \mathsf{\tilde v}|
\mathsf{u}, \mathsf{v})$ for all $\mathsf{\tilde v}$.
\end{lemma}
\begin{IEEEproof}
  Condition 1) guarantees that $\Pr\{\varPi_{U^n}(\mathsf{u}) > 0 \}
  \rightarrow 1$. Hence $\varPi_{V^n|U^n}(\mathsf{v}| \mathsf{u})$
  and $\varPi_{\tilde V^n, V^n|U^n}(\mathsf{\tilde v}, \mathsf{v}|
  \mathsf{u})$ are well-defined with high probabilities for all
  sufficiently large $n$.
Define
\begin{align}
S_n
&\defn
\sum_{i=1}^{n} \{1_i(\mathsf{w}|W^n) -  c_{\mathsf{w}|\mathsf{u}}\} 1_i(\mathsf{u}|U^n)
  \{ 1_i(\mathsf{\tilde v},
\nonumber    \mathsf{v}|\tilde V^n, V^n) \\
    &\hspace{100pt}-\varPi_{\tilde V^n, V^n|U^n}(\mathsf{\tilde v}, \mathsf{v}| \mathsf{u}) \}.
\end{align}
For any $(u^n, v^n, \tilde v^n)$ conditioned on the event $\{ U^n = u^n, V^n = v^n, \tilde V^n =
\tilde v^n\}$, conditions 2) and 3) guarantee that the sequence $\{S_n\}$ is
a martingale. According to Hoeffding's inequality \cite[Theorem 2]{Hoeffding}, we have
\begin{equation}
\Pr \left\{ \left| \frac{S_n}{n} \right| \geq \mu \bigg| U^n = u^n, V^n = v^n, \tilde V^n =
\tilde v^n \right\} \leq 2e^{-\frac{n\mu^2}{2}}
\label{e:hoeffding}
\end{equation}
for any $\mu >0$.

Furthermore, we have the equality
\begin{equation}
\sum_{i=1}^{n} 1_{i}(\mathsf{u}|U^n) \{ 1_{i}(\mathsf{\tilde v},
  \mathsf{v}|\tilde V^n, V^n) -\varPi_{\tilde V^n, V^n|U^n}
  (\mathsf{\tilde v}, \mathsf{v}| \mathsf{u}) \}= 0,
\end{equation}
which implies that
\begin{align*}
\frac{S_n}{n}
\nonumber& =  \varPi_{\tilde V^n, V^n,U^n,W^n}(\mathsf{\tilde
    v}, \mathsf{v}, \mathsf{u}, \mathsf{w})\\
  &\hspace{50pt}-\varPi_{\tilde V^n, V^n|U^n}(\mathsf{\tilde v}, \mathsf{v}| \mathsf{u}) \varPi_{U^n, W^n}(
  \mathsf{u}, \mathsf{w}).
\end{align*}
The above equation together with \eqref{e:hoeffding} gives
\begin{align}
\Pr\Big\{\nonumber&
  \big| \varPi_{\tilde V^n, V^n,U^n,W^n}(\mathsf{\tilde v}, \mathsf{v}, \mathsf{u}, \mathsf{w})\\
    &- \varPi_{\tilde V^n, V^n|U^n}(\mathsf{\tilde v}, \mathsf{v}| \mathsf{u})\varPi_{U^n,
      W^n}(\mathsf{u}, \mathsf{w}) \big| \geq \mu \Big\}
 \leq 2e^{-\frac{n\mu^2}{2}}.
\label{e:mi2}
\end{align}
Thus, we have
\begin{equation}
\label{e:5.1}
\varPi_{\tilde V^n\!,\! V^n,U^n,W^n}(\mathsf{\tilde v}, \mathsf{v}, \mathsf{u}, \mathsf{w})
 \!\!\rightarrow\!\! \varPi_{\tilde V^n, V^n|U^n}(\mathsf{\tilde v}, \mathsf{v}| \mathsf{u})
\varPi_{U^n, W^n}(\mathsf{u}, \mathsf{w})
\end{equation}
and
\begin{align}
\label{e:5.2}
\varPi_{U^n,V^n,W^n}(\mathsf{u}, \mathsf{v}, \mathsf{w})
&\rightarrow \varPi_{V^n|U^n}(\mathsf{v}| \mathsf{u}) \varPi_{U^n, W^n}(
\mathsf{u}, \mathsf{w}) \notag \\
& = \varPi_{U^n,V^n}(\mathsf{u}, \mathsf{v}) \varPi_{W^n| U^n}(
\mathsf{w}| \mathsf{u}).
\end{align}
Immediately, \eqref{e:5.2} implies $\varPi_{W^n| U^n,V^n}(\mathsf{w}|
\mathsf{u}, \mathsf{v}) \rightarrow \varPi_{W^n| U^n}(\mathsf{w}|
\mathsf{u})$. Furthermore, note that if condition 4) holds,
\eqref{e:5.2} also implies that $\Pr\{
\varPi_{U^n,V^n,W^n}(\mathsf{u}, \mathsf{v}, \mathsf{w}) > 0\}
\rightarrow 1$, and hence $\varPi_{\tilde V^n|
  U^n,V^n,W^n}(\mathsf{\tilde v}| \mathsf{u}, \mathsf{v}, \mathsf{w})$
is well-defined. Dividing \eqref{e:5.1} by \eqref{e:5.2} gives
$\varPi_{\tilde V^n| U^n,V^n,W^n}(\mathsf{\tilde v}| \mathsf{u},
\mathsf{v}, \mathsf{w}) \rightarrow \varPi_{\tilde V^n| U^n,V^n}(
\mathsf{\tilde v}| \mathsf{u}, \mathsf{v})$.
\end{IEEEproof}

Let us now turn our attention to the proof of Proposition \ref{memory_attack}. Note that
\begin{align}
\label{prop_11}
\nonumber&\varPi_{V_{1}^{n},V_{2}^{n}|U_{1}^{n},U_{2}^{n}}(\mathsf{v}_{1},\mathsf{v}_{2}|\mathsf{u}_{1},\mathsf{u}_{2})\\
&=\varPi_{V_{1}^{n}|U_{1}^{n},U_{2}^{n}}(\mathsf{v}_{1}|\mathsf{u}_{1},\mathsf{u}_{2})\varPi_{V_{2}^{n}|V_{1}^{n},U_{1}^{n},U_{2}^{n}}(\mathsf{v}_{2}|\mathsf{v}_{1},\mathsf{u}_{1},\mathsf{u}_{2}).
\end{align}Since $P_{U_1,U_2}(\mathsf{u}_1, \mathsf{u}_2) > 0$,
  we have $\Pr\{\varPi_{U_1^n,U_2^n} (\mathsf{u}_1, \mathsf{u}_2) >0\}
  \rightarrow 1$. As a result, relying on Lemma~\ref{lem5}, we obtain
  $\varPi_{V_1^n|U_1^n,U_2^n}(\mathsf{v}_1 | \mathsf{u}_1,
  \mathsf{u}_2) \rightarrow \varPi_{V_1^n|U_1^n}(\mathsf{v}_1 |
  \mathsf{u}_1)$. This result, together with the assumption $\Pr\{
  \varPi_{V_1^n|U_1^n}(\mathsf{v}_1 | \mathsf{u}_1) > 0 \} \rightarrow
  1$, gives $\Pr\{ \varPi_{V_1^n, U_1^n, U_2^n}(\mathsf{v}_1,
  \mathsf{u}_1, \mathsf{u}_2) > 0 \} \rightarrow 1$, and thus
  $\varPi_{V_2^n | V_1^n, U_1^n, U_2^n}(\mathsf{v}_2 | \mathsf{v}_1,
  \mathsf{u}_1, \mathsf{u}_2)$ is well-defined. Then, the proof of Proposition \ref{memory_attack} is equivalent to
proving
\begin{equation}
\label{prop_12}
\varPi_{V_{2}^{n}|V_{1}^{n},U_{1}^{n},U_{2}^{n}}(\mathsf{v}_{2}|\mathsf{v}_{1},\mathsf{u}_{1},\mathsf{u}_{2})\rightarrow\varPi_{V_{2}^{n}|U_{2}^{n}}(\mathsf{v}_{2}|\mathsf{u}_{2}) \end{equation} as long as the conditions stated in Proposition \ref{memory_attack}
are satisfied.
To this end, we notice that Proposition \ref{memory_attack} essentially assumes
\begin{enumerate}
\item $(V_1^n,U_1^n)$ and $V_2^n$ are conditionally independent given
  $U_2^n$.
\item There exists a constant $c_{\mathsf{v_2}|\mathsf{u_2}}$ such that
  $E\left\{ 1_i(\mathsf{v}_{2}|V_{2}^n) | U_{2}^n = u_{2}^n\right\} = c_{\mathsf{v}_2|\mathsf{u}_2}$
  for all $u_{2,i} = \mathsf{u}_{2}$.
\end{enumerate}Substituting these conditions into Lemma 1, it is straightforward to arrive at (\ref{prop_12}).
Therefore,  Proposition \ref{memory_attack} is proved.
\subsection{Proof of Proposition \ref{permutation}}
The proof of Proposition \ref{permutation} is presented as follows.
\begin{IEEEproof}
Let us consider the following two cases:
\begin{enumerate}
  \item[i)]The relays attack their received signals by rearranging the sequences in the time domain.
  For any $k,t=1,\ldots n$, if $V_{1,k}=U_{1,t}$ and $V_{1,t}=U_{1,k}$, there must be $V_{2,k}=U_{2,t}$ and $V_{2,t}=U_{2,k}$. In such case, we obtain $\varPi'_{V_{m}^{n}|U_{m}^{n}}\neq I$ for $m=1,2$.
  \item[ii)] Both relays are non-malicious, i.e., $\varPi_{V_{m}^{n}|U_{m}^{n}}=I$ for $m=1,2$.
\end{enumerate}
Let us stack $U_1$ and $U_2$ into a vector denoted as $\boldsymbol{U}$, i.e., $\boldsymbol{U}=[U_1;U_2]$.
Similarly, we also stack $V_1$ and $V_2$ into a vector denoted as $\boldsymbol{V}$, i.e., $\boldsymbol{V}=[V_1;V_2]$.
The matrices $\boldsymbol{U}^n=[U_1^n;U_2^n]$ and $\boldsymbol{V}^n=[V_1^n;V_2^n]$, of dimension $2\times n$,
represent the input and output random sequences of the relays, respectively. 
The $i$th column in 
$\boldsymbol{U}^n$ and $\boldsymbol{V}^n$ is denoted as $\boldsymbol{U}_{i}$ and $\boldsymbol{V}_{i}$, respectively.
In the case i),
from $V_{1,k}=U_{1,t},V_{1,t}=U_{1,k}$ and $V_{2,k}=U_{2,t},V_{2,t}=U_{2,k}$, we have $\boldsymbol{V}_{k}=\boldsymbol{U}_{t}$ and $\boldsymbol{V}_{t}=\boldsymbol{U}_{k}$,
which indicates that $\boldsymbol{V}^n$ is actually equivalent to the rearrangement of $\boldsymbol{U}^n$ in the time domain.
Since $\boldsymbol{U}^n$ is an i.i.d sequence, the PMF of $\boldsymbol{U}^n$ only depends on the PMFs of its elements. Rearranging $\boldsymbol{U}^n$ does not change
the PMFs of its elements, and hence the PMF of $\boldsymbol{U}^n$ remains unchanged. In other words, $\boldsymbol{V}^n$ and $\boldsymbol{U}^n$  have the same PMF in the case i).
Therefore, the distribution of $Y^n$ in
both cases is exactly the same. Thus, any decision statistic $D(Y^n)$ also
has the same distribution in both cases. Furthermore, in
case ii), since we have $\sum_{m=1}^{2} \|\varPi_{V_{m}^{n}|U_{m}^{n}} - I\|_2 =
0$, Property 2) of Theorem~\ref{thm:main} requires the probability of
the event $\{ D\left({Y}^{n}\right) > \delta\}$ be arbitrarily small as long as $\delta$
is sufficiently small and $n$ is sufficiently large. On the other hand, in the case
i), by choosing $\delta < \sum_{m=1}^{2} \|\varPi'_{V_{m}^{n}|U_{m}^{n}} -
I\|_2$, Property 1) requires that the probability of the same event
$\{ D\left({Y}^{n}\right) > \delta\}$ is arbitrarily close to $1$ as long as $n$ is
large enough. Hence, these two requirements contradict. Therefore, Proposition \ref{permutation} is proved.
\end{IEEEproof}

\section{Proof of Theorem \ref{thm:main} } \label{se:proof}
\subsection{Proof of Sufficiency of Theorem \ref{thm:main}}\label{se:suffproof}
To prove the sufficiency of Theorem 1, we need the following Lemma 2 which characterizes the convergence
property of $\varPi_{Y^n}$. Note that the lemma holds for any arbitrary attack in the family (7).
\begin{lemma}\label{lem11} If {\small{$ \varPi_{V_{1}^{n},V_{2}^{n}\left|U_{1}^{n},U_{2}^{n}\right.}\left(\mathsf{v}_{1},\mathsf{v}_{2}\left|\mathsf{u}_{1},\mathsf{u}_{2}\right.\right)\rightarrow\varPi_{V_{1}^{n}\left|U_{1}^{n}\right.}\left(\mathsf{v}_{1}\left|\mathsf{u}_{1}\right.\right)\varPi_{V_{2}^{n}\left|U_{2}^{n}\right.}\left(\mathsf{v}_{2}\left|\mathsf{u}_{2}\right.\right)$}}, we have {\small{$\big\| \varPi_{Y^{n}}-P_{{U_{1},U_{2}}}(\varPi_{{V_{1}^{n}}\left|{U_{1}^{n}}\right.}^{T}\otimes\varPi_{{V_{2}^{n}}\left|{U_{2}^{n}}\right.}^{T})P_{Y\left|{V_{1},V_{2}}\right.}^{T}\big\|_{2}
  \rightarrow 0$}} in probability as $n \rightarrow \infty$.
\end{lemma}
\begin{IEEEproof}
{According to the definitions of {\small{$\varPi_{V_{1}^{n},V_{2}^{n}\left|U_{1}^{n},U_{2}^{n}\right.}$}}, {\small{$\varPi_{V_{1}^{n}\left|U_{1}^{n}\right.}$}} and {\small{$\varPi_{V_{2}^{n}\left|U_{2}^{n}\right.}$}}, the convergence {\small{$$\varPi_{V_{1}^{n},V_{2}^{n}\left|U_{1}^{n},U_{2}^{n}\right.}\left(\mathsf{v}_{1},\mathsf{v}_{2}\left|\mathsf{u}_{1},\mathsf{u}_{2}\right.\right)\rightarrow\varPi_{V_{1}^{n}\left|U_{1}^{n}\right.}\left(\mathsf{v}_{1}\left|\mathsf{u}_{1}\right.\right)\varPi_{V_{2}^{n}\left|U_{2}^{n}\right.}\left(\mathsf{v}_{2}\left|\mathsf{u}_{2}\right.\right)$$}}indicates $$\left[\varPi_{V_{1}^{n},V_{2}^{n}\left|U_{1}^{n},U_{2}^{n}\right.}\right]_{i.j}\rightarrow\left[\varPi_{V_{1}^{n},\left|U_{1}^{n}\right.}\right]_{t_{1},k_{1}}\left[\varPi_{V_{2}^{n},\left|U_{2}^{n}\right.}\right]_{t_{2},k_{2}},$$
$$\hspace{10pt}j=\left(k_{1}-1\right)\left|\mathcal{U}_{2}\right|+k_{2},i=\left(t_{1}-1\right)\left|\mathcal{U}_{2}\right|+t_{2}.$$ Then, applying the definition of Kronecker product, we obtain {\small{$\left[\varPi_{V_{1}^{n},V_{2}^{n}\left|U_{1}^{n},U_{2}^{n}\right.}\right]_{i.j}\rightarrow\left[\varPi_{V_{1}^{n},\left|U_{1}^{n}\right.}\otimes\varPi_{V_{2}^{n},\left|U_{2}^{n}\right.}\right]_{i,j}$}}, which yields $\varPi_{{V_1^n,V_2^n}\left|{U_1^n,U_2^n}\right.}\rightarrow\varPi_{V_{1}^{n}\left|U_{1}^{n}\right.}\otimes\varPi_{V_{2}^{n}\left|U_{2}^{n}\right.}.$}

For any $\mu_1>0$, it is clear that (\ref{e:omegabound}) holds true, which is given in the next page.
\begin{figure*}[!ht]
\hrulefill
\normalsize
{{\small
\begin{align}
&\Pr\left\{ \left\| P_{{U_{1},U_{2}}}(\varPi_{{V_{1}^{n}}\left|{U_{1}^{n}}\right.}^{T}\otimes\varPi_{{V_{2}^{n}}\left|{U_{2}^{n}}\right.}^{T})P_{Y\left|{V_{1},V_{2}}\right.}^{T}
  - \varPi_{Y^n} \right\|_{2}> \mu_1 \right\}<\Pr\left\{
  \abs{P_{{U_{1},U_{2}}}(\varPi_{{V_{1}^{n}}\left|{U_{1}^{n}}\right.}^{T}\otimes\varPi_{{V_{2}^{n}}\left|{U_{2}^{n}}\right.}^{T})P_{Y\left|{V_{1},V_{2}}\right.}^{T}
  - \varPi_{Y^n}} > \mu_1 \right\}
\nonumber \\&
\leq\Pr\left\{ \abs{P_{{U_1,U_2}}\varPi_{{V_1^n,V_2^n}\left|{U_1^n,U_2^n}\right.}^{T}P_{Y\left|{V_1,V_2}\right.}^{T} -\varPi_{Y^n}} > \frac{\mu_{1}}{2} \right\}
+\nonumber \\
&\hspace{80pt}
\Pr\left\{ \abs{P_{{U_1,U_2}}\varPi_{{V_1^n,V_2^n}\left|{U_1^n,U_2^n}\right.}^{T}P_{Y\left|{V_1,V_2}\right.}^{T}- P_{{U_{1},U_{2}}}(\varPi_{{V_{1}^{n}}\left|{U_{1}^{n}}\right.}^{T}\otimes\varPi_{{V_{2}^{n}}\left|{U_{2}^{n}}\right.}^{T})P_{Y\left|{V_{1},V_{2}}\right.}^{T}
 } > \frac{\mu_1}{2} \right\}.
\label{e:omegabound}
\end{align}}}
\end{figure*}The second probability after 
the last ``$\leq$" of (\ref{e:omegabound}) approaches zero according to the above-obtained assertion that $\varPi_{{V_1^n,V_2^n}\left|{U_1^n,U_2^n}\right.}\rightarrow\varPi_{V_{1}^{n}\left|U_{1}^{n}\right.}\otimes\varPi_{V_{2}^{n}\left|U_{2}^{n}\right.}$. The first probability after the last ``$\leq$" of (\ref{e:omegabound}) also approaches zero, as shown by the following proof. For ease of description, we employ the vector notation in the following proof. For instance, the pair of random variables $\{U_1,U_2\}$ is denoted as $\boldsymbol{U}=[U_1;U_2]$, whose alphabet $\boldsymbol{\mathcal{U}}$ is the Cartesian product of $\mathcal{U}_1$ and $\mathcal{U}_2$. For ease of description,  
we also employ $\boldsymbol{\mathsf{u}}_{i}$ to denote the $i$th element in $\boldsymbol{\mathcal{U}}$, where $i$ is an integer taking value from $1$ to $|\boldsymbol{\mathcal{U}}|$. Similarly, we also stack $V_1$ and $V_2$ into a vector denoted as $\boldsymbol{V}$, i.e., $\boldsymbol{V}=[V_1;V_2]$, whose alphabet $\boldsymbol{\mathcal{V}}$ is the Cartesian product of $\mathcal{V}_1$ and $\mathcal{V}_2$. For convenience of exposition, 
we employ $\boldsymbol{\mathsf{v}}_{i}$ to denote the $i$th element in $\boldsymbol{\mathcal{V}}$, where $i$ is an integer taking value from $1$ to $|\boldsymbol{\mathcal{V}}|$. 
Again, $\boldsymbol{U}^n=[U_1^n;U_2^n]$ and $\boldsymbol{V}^n=[V_1^n;V_2^n]$
denote the input and output random sequences of the relays, respectively. 
Correspondingly, $\boldsymbol{u}^n=[u_1^n;u_2^n]$ and $\boldsymbol{v}^n=[v_1^n;v_2^n]$
denote the generic value of the input and output sequences of the relays, respectively. 
In other words, $\boldsymbol{U}^n$ and $\boldsymbol{V}^n$ are sequences
of random variables, while $\boldsymbol{u}^n$ and $\boldsymbol{v}^n$ denote
the possible value of $\boldsymbol{U}^n$ and $\boldsymbol{V}^n$, respectively.
Then, we have
{\small{\begin{align}
\nonumber  \abs{ P_{\boldsymbol{U}}\varPi_{\boldsymbol{V}^{n}\left|\boldsymbol{U}^{n}\right.}^{T}P_{Y\left|\boldsymbol{V}\right.}^{T}-\varPi_{Y^{n}}} &=\sum_{i=1}^{\left|\mathcal{Y}\right|}\abs{ [P_{\boldsymbol{U}}\varPi_{\boldsymbol{V}^{n}\left|\boldsymbol{U}^{n}\right.}^{T}P_{Y\left|\boldsymbol{V}\right.}^{T}]_{i}-[\varPi_{Y^{n}}]_{i}} \\&\leq\sum_{i=1}^{\left|\mathcal{Y}\right|}\sum_{j=1}^{\left|\boldsymbol{\mathcal{U}}\right|}\sum_{k=1}^{\left|\boldsymbol{\mathcal{V}}\right|}\abs{ H_{i,j,k}},
 \end{align}}}where
{\small{\begin{align}
\nonumber&H_{i,j,k}=P_{\boldsymbol{U}}\left(\boldsymbol{\mathsf{u}}_{j}\right)\frac{N\left(\boldsymbol{\mathsf{u}}_{j},\boldsymbol{\mathsf{v}}_{k}\left|\boldsymbol{U}^{n},\boldsymbol{V}^{n}\right.\right)}{N\left(\boldsymbol{\mathsf{u}}_{j}\left|\boldsymbol{U}^{n}\right.\right)}P_{Y\left|\boldsymbol{V}\right.}\left(\mathsf{y}_{i}\left|\boldsymbol{\mathsf{v}}_{k}\right.\right)\\
&\hspace{100pt}-\frac{N\left(\boldsymbol{\mathsf{u}}_{j},\boldsymbol{\mathsf{v}}_{k},\mathsf{y}_{i}\left|\boldsymbol{U}^{n},\boldsymbol{V}^{n},Y^{n}\right.\right)}{n}.
 \end{align}}}This
implies that
\begin{align}
\nonumber&\Pr\left\{\abs{ P_{\boldsymbol{U}}\varPi_{\boldsymbol{V}^{n}\left|\boldsymbol{U}^{n}\right.}^{T}P_{Y\left|\boldsymbol{V}\right.}^{T}-\varPi_{Y^{n}}} >\frac{\mu_{1}}{2}\right\}\\
&\hspace{20pt}\leq\sum_{i=1}^{\left|\mathcal{Y}\right|}\sum_{j=1}^{\left|\boldsymbol{\mathcal{U}}\right|}\sum_{k=1}^{\left|\boldsymbol{\mathcal{V}}\right|}\Pr\left\{ \abs{ H_{i,j,k}} \geq\frac{\mu_{1}}{2\left|\mathcal{Y}\right|\left|\boldsymbol{\mathcal{U}}\right|\left|\boldsymbol{\mathcal{V}}\right|}\right\} .
  \label{e:omegabound1}
\end{align}In order to bound $\Pr\left\{ \left\Vert H_{i,j,k}\right\Vert \geq\frac{\mu_{1}}{2\left|\mathcal{Y}\right|\left|\boldsymbol{\mathcal{U}}\right|\left|\boldsymbol{\mathcal{V}}\right|}\right\}$, we first consider to bound
\begin{equation}
\Pr\left\{ \abs{\widetilde{H}_{i,j,k}} \geq\frac{\mu_{1}}{2\left|\mathcal{Y}\right|\left|\boldsymbol{\mathcal{U}}\right|\left|\boldsymbol{\mathcal{V}}\right|}\right\},
\end{equation}where
\begin{align}
\nonumber&\widetilde{H}_{i,j,k}=\frac{N\left(\boldsymbol{\mathsf{u}}_{j}\left|\boldsymbol{U}^{n}\right.\right)}{n}\frac{N\left(\boldsymbol{\mathsf{u}}_{j},\boldsymbol{\mathsf{v}}_{k}\left|\boldsymbol{U}^{n},\boldsymbol{V}^{n}\right.\right)}{N\left(\boldsymbol{\mathsf{u}}_{j}\left|\boldsymbol{U}^{n}\right.\right)}P_{Y\left|\boldsymbol{V}\right.}\left(\mathsf{y}_{i}\left|\boldsymbol{\mathsf{v}}_{k}\right.\right)\\
\nonumber&\hspace{80pt}-\frac{N\left(\boldsymbol{\mathsf{u}}_{j},\boldsymbol{\mathsf{v}}_{k},\mathsf{y}_{i}\left|\boldsymbol{U}^{n},\boldsymbol{V}^{n},Y^{n}\right.\right)}{n}\\
\nonumber&=\frac{N\left(\boldsymbol{\mathsf{u}}_{j},\boldsymbol{\mathsf{v}}_{k}\left|\boldsymbol{U}^{n},\boldsymbol{V}^{n}\right.\right)}{n}P_{Y\left|\boldsymbol{V}\right.}\left(\mathsf{y}_{i}\left|\boldsymbol{\mathsf{v}}_{k}\right.\right)\\
&\hspace{80pt}-\frac{N\left(\boldsymbol{\mathsf{u}}_{j},\boldsymbol{\mathsf{v}}_{k},\mathsf{y}_{i}\left|\boldsymbol{U}^{n},\boldsymbol{V}^{n},Y^{n}\right.\right)}{n}.
 \end{align}
For any $i$, $j$ and $k$,
{\small{\begin{equation}
\Pr\left\{\abs{ \widetilde{H}_{i,j,k}} \geq\frac{\mu_{1}}{2\left|\mathcal{Y}\right|\left|\boldsymbol{\mathcal{U}}\right|\left|\boldsymbol{\mathcal{V}}\right|}\right\}\leq\frac{4\left|\mathcal{Y}\right|^{2}\left|\boldsymbol{\mathcal{U}}\right|^{2}\left|\boldsymbol{\mathcal{V}}\right|^{2}}{\mu^{2}}E\{\abs{\widetilde{H}_{i,j,k}} ^{2}\}.
\end{equation}}}Furthermore, we have (\ref{hh_bound}), as given on the next page,
\begin{figure*}[!ht]
\normalsize
{\small{\begin{align}
\label{hh_bound}
\nonumber &E\{\abs{\widetilde{H}_{i,j,k}}^{2}\}=
E\big\{\bigg|\frac{N\left(\boldsymbol{\mathsf{u}}_{j},\boldsymbol{\mathsf{v}}_{k}\left|\boldsymbol{U}^{n},\boldsymbol{V}^{n}\right.\right)}{n}P_{Y\left|\boldsymbol{V}\right.}\left(\mathsf{y}_{i}\left|\boldsymbol{\mathsf{v}}_{k}\right.\right)-
\frac{N\left(\boldsymbol{\mathsf{u}}_{j},\boldsymbol{\mathsf{v}}_{k},\mathsf{y}_{i}\left|\boldsymbol{U}^{n},\boldsymbol{V}^{n},Y^{n}\right.\right)}{n}\bigg|^{2}\big\}
\\
\nonumber &\leq\frac{E\{\left(\sum_{t=1}^{n}\left(P_{Y\left|\boldsymbol{V}\right.}\left(\mathsf{y}_{i}\left|\boldsymbol{\mathsf{v}}_{k}\right.\right)1_{t}\left(\boldsymbol{\mathsf{u}}_{j},\boldsymbol{\mathsf{v}}_{k}\right)-1_{t}\left(\boldsymbol{\mathsf{u}}_{j},\boldsymbol{\mathsf{v}}_{k},\mathsf{y}_{i}\right)\right)\right)^{2}\}}{n^{2}}
\\
\nonumber &=\frac{E\{\left(\sum_{t=1}^{n}1_{t}\left(\boldsymbol{\mathsf{u}}_{j},\boldsymbol{\mathsf{v}}_{k}\right)\left(P_{Y\left|\boldsymbol{V}\right.}\left(\mathsf{y}_{i}\left|\boldsymbol{\mathsf{v}}_{k}\right.\right)-1_{t}\left(\mathsf{y}_{i}\right)\right)\right)^{2}\}}{n^{2}}
\\
\nonumber &=\frac{E\{
\underset{t,t'=1}{\overset{n}{\sum}}1_{t}\left(\boldsymbol{\mathsf{u}}_{j},\boldsymbol{\mathsf{v}}_{k}\right)1_{t'}\left(\boldsymbol{\mathsf{u}}_{j},\boldsymbol{\mathsf{v}}_{k}\right)\left(P_{Y\left|\boldsymbol{V}\right.}\left(\mathsf{y}_{i}\left|\boldsymbol{\mathsf{v}}_{k}\right.\right)-1_{t}\left(\mathsf{y}_{i}\right)\right)
\left(P_{Y\left|\boldsymbol{V}\right.}\left(\mathsf{y}_{i}\left|\boldsymbol{\mathsf{v}}_{k}\right.\right)-1_{t'}\left(\mathsf{y}_{i}\right)\right)
\}}{n^{2}}
\\
\nonumber &\leq\frac{E\{\underset{t=1}{\overset{n}{\sum}}1_{t}\left(\boldsymbol{\mathsf{u}}_{j},\boldsymbol{\mathsf{v}}_{k}\right)\left(P_{Y\left|\boldsymbol{V}\right.}\left(\mathsf{y}_{i}\left|\boldsymbol{\mathsf{v}}_{k}\right.\right)-1_{t}\left(\mathsf{y}_{i}\right)\right)^{2}\}}{n^{2}}
\\\nonumber &\hspace{70pt}+\frac{E\{\underset{t,t'=1,t\neq t'}{\overset{n}{\sum}}1_{t}\left(\boldsymbol{\mathsf{u}}_{j},\boldsymbol{\mathsf{v}}_{k}\right)1_{t'}\left(\boldsymbol{\mathsf{u}}_{j},\boldsymbol{\mathsf{v}}_{k}\right)\left(P_{Y\left|\boldsymbol{V}\right.}\left(\mathsf{y}_{i}\left|\boldsymbol{\mathsf{v}}_{k}\right.\right)-1_{t}\left(\mathsf{y}_{i}\right)\right)\left(P_{Y\left|\boldsymbol{V}\right.}\left(\mathsf{y}_{i}\left|\boldsymbol{\mathsf{v}}_{k}\right.\right)-1_{t'}\left(\mathsf{y}_{i}\right)\right)\}}{n^{2}}
\\
\nonumber &\overset{\left(a\right)}{\leq}\frac{E_{\boldsymbol{U}^{n},\boldsymbol{V}^{n}}\left\{\underset{t=1}{\overset{n}{\sum}}E\left\{1_{t}\left(\boldsymbol{\mathsf{u}}_{j},\boldsymbol{\mathsf{v}}_{k}\right)\left(P_{Y\left|\boldsymbol{V}\right.}\left(\mathsf{y}_{i}\left|\boldsymbol{\mathsf{v}}_{k}\right.\right)-1_{t}\left(\mathsf{y}_{i}\right)\right)^{2}\left|\boldsymbol{u}^{n},\boldsymbol{v}^{n}\right.\right\}\right\}}{n^{2}}
\\
\nonumber&\overset{\left(b\right)}{\leq}\frac{\underset{t=1}{\overset{n}{\sum}}E\left\{\left(P_{Y\left|\boldsymbol{V}\right.}\left(\mathsf{y}_{i}\left|\boldsymbol{\mathsf{v}}_{k}\right.\right)-1_{t}\left(\mathsf{y}_{i}\right)\right)^{2}\left|\boldsymbol{\mathsf{u}}_{j},\boldsymbol{\mathsf{v}}_{k}\right.\right\}}{n^{2}}
\\
&\leq\frac{P_{Y\left|\boldsymbol{V}\right.}^{2}\left(\mathsf{y}_{i}\left|\boldsymbol{\mathsf{v}}_{k}\right.\right)}{n},
\end{align}}}
\end{figure*}
where the inequality (b) is obtained based on the fact that the elements of $Y^{n}$ are conditionally independent given $\boldsymbol{V}^{n}=\boldsymbol{v}^{n}$. Again relying on this fact, the inequality (a) is obtained as follows. Firstly, we have (\ref{e}), as given on the next page.
\begin{figure*}[!ht]
\normalsize
{{\small{\begin{align}\label{e}
\nonumber&E\left\{\underset{t,t'=1,t\neq t'}{\overset{n}{\sum}}1_{t}\left(\boldsymbol{\mathsf{u}}_{j},\boldsymbol{\mathsf{v}}_{k}\right)1_{t'}\left(\boldsymbol{\mathsf{u}}_{j},\boldsymbol{\mathsf{v}}_{k}\right)\left(P_{Y\left|\boldsymbol{V}\right.}\left(\mathsf{y}_{i}\left|\boldsymbol{\mathsf{v}}_{k}\right.\right)-1_{t}\left(\mathsf{y}_{i}\right)\right)\left(P_{Y\left|\boldsymbol{V}\right.}\left(\mathsf{y}_{i}\left|\boldsymbol{\mathsf{v}}_{k}\right.\right)-1_{t'}\left(\mathsf{y}_{i}\right)\right)\right\}\\&=
E_{\boldsymbol{U}^{n},\boldsymbol{V}^{n}}\left\{\underset{t,t'=1,t\neq t'}{\overset{n}{\sum}}E\left\{1_{t}\left(\boldsymbol{\mathsf{u}}_{j},\boldsymbol{\mathsf{v}}_{k}\right)1_{t'}\left(\boldsymbol{\mathsf{u}}_{j},\boldsymbol{\mathsf{v}}_{k}\right)\left(P_{Y\left|\boldsymbol{V}\right.}\left(\mathsf{y}_{i}\left|\boldsymbol{\mathsf{v}}_{k}\right.\right)-1_{t}\left(\mathsf{y}_{i}\right)\right)\left(P_{Y\left|\boldsymbol{V}\right.}\left(\mathsf{y}_{i}\left|\boldsymbol{\mathsf{v}}_{k}\right.\right)-1_{t'}\left(\mathsf{y}_{i}\right)\right)\left|\boldsymbol{u}^{n},\boldsymbol{v}^{n}\right.\right\}\right\}.
\end{align}}}}
\end{figure*}Furthermore, for each $\left(\boldsymbol{u}^{n},\boldsymbol{v}^{n}\right)$ and $\left(t,t'\right)$ with $1_{t}\left(\boldsymbol{\mathsf{u}}_{j},\boldsymbol{\mathsf{v}}_{k}\right)1_{t'}\left(\boldsymbol{\mathsf{u}}_{j},\boldsymbol{\mathsf{v}}_{k}\right)=1$, we have (\ref{e12}) shown on the next page,
\begin{figure*}[!ht]
\normalsize
{{\small{\begin{align}\label{e12}
\nonumber&E\left\{1_{t}\left(\boldsymbol{\mathsf{u}}_{j},\boldsymbol{\mathsf{v}}_{k}\right)1_{t'}\left(\boldsymbol{\mathsf{u}}_{j},\boldsymbol{\mathsf{v}}_{k}\right)\left(P_{Y\left|\boldsymbol{V}\right.}\left(\mathsf{y}_{i}\left|\boldsymbol{\mathsf{v}}_{k}\right.\right)-1_{t}\left(\mathsf{y}_{i}\right)\right)\left(P_{Y\left|\boldsymbol{V}\right.}\left(\mathsf{y}_{i}\left|\boldsymbol{\mathsf{v}}_{k}\right.\right)-1_{t'}\left(\mathsf{y}_{i}\right)\right)\left|\boldsymbol{u}^{n},\boldsymbol{v}^{n}\right.\right\}\\
\nonumber&=E\left\{\left(P_{Y\left|\boldsymbol{V}\right.}\left(\mathsf{y}_{i}\left|\boldsymbol{\mathsf{v}}_{k}\right.\right)-1_{t}\left(\mathsf{y}_{i}\right)\right)\left(P_{Y\left|\boldsymbol{V}\right.}\left(\mathsf{y}_{i}\left|\boldsymbol{\mathsf{v}}_{k}\right.\right)-1_{t'}\left(\mathsf{y}_{i}\right)\right)\left|1_{t}\left(\boldsymbol{\mathsf{u}}_{j},\boldsymbol{\mathsf{v}}_{k}\right)1_{t'}\left(\boldsymbol{\mathsf{u}}_{j},\boldsymbol{\mathsf{v}}_{k}\right)=1\right.\right\}\\
\nonumber&=E\left\{\left(P_{Y\left|\boldsymbol{V}\right.}\left(\mathsf{y}_{i}\left|\boldsymbol{\mathsf{v}}_{k}\right.\right)-1_{t}\left(\mathsf{y}_{i}\right)\right)\left|\boldsymbol{\mathsf{u}}_{j},\boldsymbol{\mathsf{v}}_{k}\right.\right]E\left[\left(P_{Y\left|\boldsymbol{V}\right.}\left(\mathsf{y}_{i}\left|\boldsymbol{\mathsf{v}}_{k}\right.\right)-1_{t'}\left(\mathsf{y}_{i}\right)\right)\left|\boldsymbol{\mathsf{u}}_{j},\boldsymbol{\mathsf{v}}_{k}\right.\right\}\\
\nonumber&\overset{\left(a\right)}{=}\left(P_{Y\left|\boldsymbol{V}\right.}\left(\mathsf{y}_{i}\left|\boldsymbol{\mathsf{v}}_{k}\right.\right)-P_{Y\left|\boldsymbol{V}\right.}\left(\mathsf{y}_{i}\left|\boldsymbol{\mathsf{v}}_{k}\right.\right)\right)
\left(P_{Y\left|\boldsymbol{V}\right.}\left(\mathsf{y}_{i}\left|\boldsymbol{\mathsf{v}}_{k}\right.\right)-P_{Y\left|\boldsymbol{V}\right.}\left(\mathsf{y}_{i}\left|\boldsymbol{\mathsf{v}}_{k}\right.\right)\right)\\
&=0,
\end{align}}}}
\hrulefill
\end{figure*}where the equation (a) is obtained relying on {\small{$P_{Y\left|\boldsymbol{V}\right.}\left(\mathsf{y}_{i}\left|\boldsymbol{\mathsf{v}}_{k}\right.\right)=P_{Y\left|\boldsymbol{V,U}\right.}\left(\mathsf{y}_{i}\left|\boldsymbol{\mathsf{v}}_{k},\boldsymbol{\mathsf{u}}_{j}\right.\right)$}}. Substituting (\ref{hh_bound}) into (32), we get {\small{$\Pr \{ \abs{ \widetilde{H}_{i,j,k}} \geq\frac{\mu_{1}}{2\left|\mathcal{Y}\right|\left|\boldsymbol{\mathcal{U}}\right|\left|\boldsymbol{\mathcal{V}}\right|}\} \rightarrow0$}} as {\small{$n\rightarrow\infty$}}. Comparing $\widetilde{H}_{i,j,k}$ with $H_{i,j,k}$, we see that $\widetilde{H}_{i,j,k}\rightarrow\ensuremath{H_{i,j,k}}$ as $n\rightarrow\infty$, then we arrive at {\small{$\Pr\left\{ \abs{ H_{i,j,k}} \geq\frac{\mu_{1}}{2\left|\mathcal{Y}\right|\left|\boldsymbol{\mathcal{U}}\right|\left|\boldsymbol{\mathcal{V}}\right|}\right\} \rightarrow0$ as $n\rightarrow\infty$}}. Using (\ref{e:omegabound1}), we further obtain $\Pr\left\{\abs{P_{\boldsymbol{U}}\varPi_{\boldsymbol{V}^{n}\left|\boldsymbol{U}^{n}\right.}^{T}P_{Y\left|\boldsymbol{V}\right.}^{T}-\varPi_{Y^{n}}} >\frac{\mu_{1}}{2}\right\}\rightarrow0$ as $n\rightarrow\infty$.
Thus, this lemma has been proved, because we have shown that the two probabilities after the last ``$\leq$" of (\ref{e:omegabound}) converge to $0$ as $n$
approaches infinity.
\end{IEEEproof}

Applying Lemma 2, the sufficiency proof of Theorem 1 originally outlined in \cite{cao2013detecting} for dealing with i.i.d attacks can be
readily extended to the case of non-i.i.d attacks considered here. We provide the details of proof below for completeness.

To establish the proof of sufficiency, we construct the estimators $\hat{\varPi}_{V_{1}^{n}|U_{1}^{n}}$ and
$\hat{\varPi}_{V_{2}^{n}|U_{2}^{n}}$ from $\varPi_{Y^n}$ according to the
following arrangement: \\
For $\mu >0$, let $\mathcal{G}_{\mu}(\varPi_{Y^n})$ be the set of
all pairs of the two stochastic matrices
${P}_{1}$ and ${P}_{2}$ (their dimensions are $|\mathcal{U}_{1}|\times|\mathcal{U}_{1}| $ and $|\mathcal{U}_{2}|\times|\mathcal{U}_{2}|$, respectively), which satisfy
\begin{equation}
\left\Vert P_{{U_{1},U_{2}}}({P}_{1}^{T}\otimes{P}_{2}^{T})P_{Y\left|{V_{1},V_{2}}\right.}^{T}-\varPi_{Y^n}\right\Vert _{2}\leq\mu;
\end{equation}
if  $\mathcal{G}_{\mu}(\varPi_{Y^n})$ is non-empty, we set
\begin{equation}
\label{esti}
(\hat{\varPi}_{V_{1}^{n}|U_{1}^{n}},\hat{\varPi}_{V_{2}^{n}|U_{2}^{n}})
= \hspace*{-15pt}
\underset{_{(P_{1},P_{2})\in
      \mathcal{G}_{\mu}(\varPi_{Y^n})}}{\arg\max}
\hspace*{-0pt} \sum_{i=1}^{2}\|P_{i}-I\|_{2};
\end{equation}
otherwise, set $(\hat{\varPi}_{V_{1}^{n}|U_{1}^{n}},\hat{\varPi}_{V_{2}^{n}|U_{2}^{n}}) = (I,I)$. Relying on
$(\hat{\varPi}_{V_{1}^{n}|U_{1}^{n}},\hat{\varPi}_{V_{2}^{n}|U_{2}^{n}})$, in what follows
we employ the decision statistic $D\left({Y}^{n}\right) = \sum_{i=1}^{2} \|
\hat{\varPi}_{V_{i}^{n}|U_{i}^{n}}- I \|_2$.
\subsubsection{The Proof of Property 1) of Theorem 1}
To show that this decision statistic satisfies
Property 1) of Theorem 1, note that (\ref{proof1}) holds,
\begin{figure*}[!ht]
\hrulefill
\begin{align} \label{proof1}
\nonumber &\Pr\left\{
  D\left({Y}^{n}\right)>\delta ~\bigcap~
  \sum_{i=1}^{2} \|{\varPi}_{V_{i}^{n}|U_{i}^{n}}-I\|_{2} >\delta \right\} \\
\nonumber &\geq \Pr\bigg\{
  ({\varPi}_{V_{1}^{n}|U_{1}^{n}},{\varPi}_{V_{2}^{n}|U_{2}^{n}}) \in
  \mathcal{G}_{\mu}(\varPi_{Y^n}) ~\bigcap~ D\left({Y}^{n}\right) > \delta \bigcap~
  \sum_{i=1}^{2} \|{\varPi}_{V_{i}^{n}|U_{i}^{n}}-I\|_{2} >\delta  \bigg\}\\
\nonumber & = \Pr\bigg\{
 ({\varPi}_{V_{1}^{n}|U_{1}^{n}},{\varPi}_{V_{2}^{n}|U_{2}^{n}}) \in
  \mathcal{G}_{\mu}(\varPi_{Y^n})
\bigcap~ \sum_{i=1}^{2} \|{\varPi}_{V_{i}^{n}|U_{i}^{n}}-I\|_{2} >\delta  \bigg\}\\
 & \geq \Pr\left\{
  \sum_{i=1}^{2} \|{\varPi}_{V_{i}^{n}|U_{i}^{n}}-I\|_{2} >\delta\right\}
 - \Pr\left\{ ({\varPi}_{V_{1}^{n}|U_{1}^{n}},{\varPi}_{V_{2}^{n}|U_{2}^{n}}) \notin
  \mathcal{G}_{\mu}(\varPi_{Y^n}) \right\},
\end{align}
\hrulefill
\end{figure*}where the equality in the third line is obtained relying on
$\hat{\varPi}_{V_{1}^{n}|U_{1}^{n}}$ and
$\hat{\varPi}_{V_{2}^{n}|U_{2}^{n}}$ given by (\ref{esti}), and on the fact that $\Pr\bigg\{
 ({\varPi}_{V_{1}^{n}|U_{1}^{n}},{\varPi}_{V_{2}^{n}|U_{2}^{n}}) \in
  \mathcal{G}_{\mu}(\varPi_{Y^n}) \bigg\}$ as
$n\rightarrow\infty$, as implied by Lemma~\ref{lem11}.  Hence, it is plausible that Property 1) of Theorem 1 is a direct consequence of this latter fact
and (\ref{proof1}).
\subsubsection{The Proof of Property 2) of Theorem 1}
To prove Property 2) of Theorem 1,
we define the function\begin{align*}
F(\mathbf{w}) \defn \left\Vert P_{{U_{1},U_{2}}}(W_{1}^{T}\otimes W_{2}^{T})P_{Y\left|{V_{1},V_{2}}\right.}^{T}-P_{{U_{1},U_{2}}}P_{Y\left|{V_{1},V_{2}}\right.}^{T}\right\Vert _{2}^{2},
\end{align*}where we use $W_{i}$, of dimension $|\mathcal{U}_{i}|\times|\mathcal{U}_{i}|$, to denote all possible values of $\varPi_{{V_{i}^{n}}\left|{U_{i}^{n}}\right.}$ for
$i=1,2$. Since $\varPi_{{V_{i}^{n}}\left|{U_{i}^{n}}\right.}$ is a stochastic matrix, $W_i$ can be expressed as
{\small{\[
W_{i}= {\small \left[\begin{array}{cccc}
1-\sum_{k=2}^{|\mathcal{U}_{i}|}\left[W_{i}\right]_{k,1} & \cdots & \cdots & \left[W_{i}\right]_{1,|\mathcal{U}_{i}|}\\
\left[W_{i}\right]_{2,1} & \ddots &  & \left[W_{i}\right]_{2,|\mathcal{U}_{i}|}\\
\vdots &  & \ddots & \vdots\\
\left[W_{i}\right]_{|\mathcal{U}_{i}|,1} & \cdots & \cdots & 1-\sum_{k=1}^{|\mathcal{U}_{i}|-1}\left[W_{i}\right]_{k,|\mathcal{U}_{i}|}
\end{array}\right]},
\]}}where each entry of {\small{$W_{i}$}} takes value from the interval {\small{$[0,1]$}}.
Then, except the diagonal entries, stacking all entries of both $W_{1}$ and $W_{2}$ column by column, we obtain the vector $\mathbf{w}$.
More precisely, {\small{$\mathbf{w}=[\left[W_{1}\right]_{2,1},\ldots\left[W_{1}\right]_{|\mathcal{U}_{1}|,1},\ldots\left[W_{1}\right]_{|\mathcal{U}_{1}|-1,|\mathcal{U}_{1}|}\ldots\left[W_{2}\right]_{|\mathcal{U}_{2}|-1,|\mathcal{U}_{2}|}]$}}.
It is straightfoward to check that $\mathbf{w}$ and
$F(\mathbf{w})$ have the following properties:\begin{enumerate}
\item All possible $\mathbf{w}$'s belong to 
  {\small{$\mathcal{D}=\big\{ \mathbf{w} |0\leq\left[W_{i}\right]_{k,j}\leq1,\sum_{k=1,k\neq j}^{|\mathcal{U}_{i}|}\left[W_{i}\right]_{k,j}\leq1,\, i=1,2,\: k\neq j,\, k,j=1,2,\ldots|\mathcal{U}_{i}|\big\} $}}. $\mathcal{D}$ is a bounded convex and continuous set. Obviously, it includes $\mathbf{w}_0 \defn [0,0,\ldots,0]$.
\item $F(\mathbf{w})$ is twice continuously differentiable in $\mathcal{D}$.
\item Since the non-manipulable condition is satisfied, $\mathbf{w}_0$, which corresponds to $W_1=I$ and $W_2=I$, is the
  unique solution to $F(\mathbf{w})=0$ in $\mathcal{D}$.  Moreover, $\nabla F =
  0$ at $\mathbf{w}_0$.
\end{enumerate}
These properties can be used for proving the following Lemma 3:
\begin{lemma} \label{thm:lem2}

  There exists an $r_{1}>0$, depending only on $F(\mathbf{w})$, such that the closed
  subset $\mathcal{D}_{1}\defn\left\{ \mathbf{w}\left|\|\mathbf{w}-\mathbf{w}_{0}\|_{2}\leq r_{1}\right.\right\} \cap\mathcal{D}$ of $\mathcal{D}$ has the following
  property: \\
  For each radial line $L$ emanating from $\mathbf{w}_{0}$ to a point on the
  boundary of $\mathcal{D}_1$ (i.e., $\left\{\mathbf{w}\left| \|\mathbf{w}-\mathbf{w}_{0}\|_2=r_{1}\right.
  \right\} \cap \mathcal{D}$), we have $\nabla_L F(\mathbf{w}) > 0$ for all $\mathbf{w}\neq \mathbf{w}_0$
  on $L$, where $\nabla_L F$ is the orientational derivative of $F$
  along $L$.
\end{lemma}
\begin{IEEEproof}
Let us choose an arbitrary orientation from $\mathbf{w}_{0}$, denoted as $L=[l_{1},l_{2},\ldots,l_{|\mathbf{w}|}]$, $||L||_{2}^{2}=1$.
Each point $\mathbf{w}$ along the orientation $L$ can be expressed as $\mathbf{w}=lL$, where $l=\|\mathbf{w}-\mathbf{w}_{0}\|_{2}$ is positive and continuous-valued.
Then, relying on the orientation $L$, $l$ could be used to represent $\mathbf{w}$. We thus rewrite $\mathbf{w}$ as $\mathbf{w}_{l}$.
Correspondingly, we define a function $f\left(l\right)\defn F\left(\mathbf{w}_{l}\right)$ and its derivative function (i.e.,$f'\left( l\right)\defn \nabla_L F(\mathbf{w})$) along the orientation $L$. 
According to the third property of $F\left(\mathbf{w}\right)$, $l=0$ is the only solution to $f\left( l \right)=0$, and we also have $f'\left(0 \right)=0$. 
To complete the proof, let us prove that there exists
a positive $r_L$ such that for any arbitrary $l\in(0,\, r_{L}]$, we always have $f'\left( l\right)>0$. 
To this end, let us discuss two possible cases in the
following. In the first case where $f'\left( l\right)\neq0$ holds true for
all $l\in\left(0,\,\infty\right)$, there must exist at least one positive $r_L$ that renders $f'\left( l\right)>0$ true for all $l\in (0,\, r_{L}]$.
Otherwise, the contrary statement that $f'\left(
  l\right)<0$ for all $l\in (0,\,r_L ]$ will result in the fact that $f\left(l\right)$ is a 
  decreasing function in the domain $(0,\, r_{L}]$.  
  As a consequence, we will have $f\left(l\right)<f\left(0\right)=0$ for $l\in (0,\, r_{L}]$, which contradicts the fact that
$f\left(0\right)=0$ is the minimum value. 
In addition, if $f'\left( l\right)<0$ holds true for some values of $l$ and
$f'\left( l\right)>0$ for another set of values of $l$, we can deduce
that there is at least one point making $f'\left( l\right)=0$, since
$f'\left( l\right)$ is a continuous function according to the second property that $F(\mathbf{w})$ is twice continuously differentiable. This in turn
contradicts the basic assumption of this case, namely $f'\left( l\right)\neq0$.

Second, if $f'\left( l\right)$ can be equal to $0$, among all the points
making $f'\left( l\right)=0$, let us choose the one nearest to $l=0$, denoted as $s$.
Then, $r_L$ can be chosen from the interval $(0, s)$, such that 
for $l \in (0, r_L]$, we have $f'\left(
  l\right)\neq0$. Hence, there must be $f'\left( l\right)>0$ for $l \in (0, r_L]$. The
explanation is similar to that of the first case.

Finally,  we choose $r_1$ according to $r_{1}=\underset{L=\frac{\mathbf{w}-\mathbf{w}_{0}}{||\mathbf{w}-\mathbf{w}_{0}||_{2}},\mathbf{w}\in \mathcal{D}}{\min}r_{L}$.
Hence, for each radial line $L$ emanating from $\mathbf{w}_0$ to the boundary of $\mathcal{D}_{1}=\left\{ \mathbf{w}\left|\|\mathbf{w}-\mathbf{w}_{0}\|_{2}\leq r_{1}\right.\right\} \cap\mathcal{D}$, we have $\nabla_L F(\mathbf{w}) > 0$ for all $\mathbf{w}\neq \mathbf{w}_0$
  on $L$.
\end{IEEEproof}
Relying on Lemma~\ref{thm:lem2}, the function $F_{\min}(r) \defn \underset{\mathbf{w}:
  \|\mathbf{w}-\mathbf{w}_{0}\|_2 = r, \mathbf{w}\in \mathcal{D}}{\min} F(\mathbf{w})$ is a strictly increasing and bounded
function over $0 \leq r \leq r_1$. Hence, an inverse of $F_{\min}$
(denoted by $F^{-1}_{\min}$) exists, which is also increasing and
bounded. This property can be used to prove the following Lemma \ref{thm:lem3}:
\begin{lemma} \label{thm:lem3}
Let us use $\mathfrak{m}$ to denote $\underset{\mathbf{w}\in\mathcal{D}-\mathcal{D}_{1}}{\min}F(\mathbf{w})$.
  For all $\mathfrak{m}'\in\left(0,\,\mathfrak{m}\right]$, if $\mathbf{w}\in\mathcal{D}$ and $F(\mathbf{w}) < \mathfrak{m}'$, then there must exist
  \begin{equation}
  \eu{\mathbf{w}-\mathbf{w}_{0}} < F^{-1}_{\min}(\mathfrak{m}').
  \label{eq:invF}
  \end{equation}
\end{lemma}
\begin{IEEEproof}
Since $\mathbf{w}_{0}$ is not in $\mathcal{D}-\mathcal{D}_{1}$ and $\mathbf{w}_{0}$
is the only point satisfying $F\left(\mathbf{w}_{0}\right)=0$, $\mathfrak{m}$ is a strictly positive number. Furthermore, noticing that $\left\{ \left.\mathbf{w}\right|\left\Vert \mathbf{w}-\mathbf{w}_{0}\right\Vert_2 =r_{1}, \mathbf{w}\in \mathcal{D}\right\}$ is a part of the boundary of $\mathcal{D}-\mathcal{D}_{1}$, we have $\mathfrak{m}\leqslant F_{\min}\left(r_{1}\right)=\underset{\mathbf{w}:\|\mathbf{w}-\mathbf{w}_{0}\|_{2}=r_{1}, \mathbf{w}\in \mathcal{D}}{\min}F(\mathbf{w})$. Then, for any arbitrary $\mathfrak{m}'$ smaller than $\mathfrak{m}$, according to the monotonically increasing property of $F_{\min}^{-1}\left(\cdot\right)$, we have $F_{\min}^{-1}\left(\mathfrak{m}'\right)<F_{\min}^{-1}\left(\mathfrak{m}\right)\leqslant r_{1}$. Based on this observation, we obtain $\mathcal{D}'\subset \mathcal{D}_{1}$, where $\mathcal{D}'$ denotes $\left\{ \mathbf{w}\left|\eu{\mathbf{w}-\mathbf{w}_{0}}<F_{\min}^{-1}\left(\mathfrak{m}'\right)\right.\right\} \cap D$. As a result, $\mathcal{D}-\mathcal{D}^{'}=\left\{ \mathcal{D}-\mathcal{D}_{1}\right\} \cup\left\{ \mathcal{D}_{1}-\mathcal{D}^{'}\right\} $ holds true, from which the minimum value of $F(\mathbf{w})$ over $\mathcal{D}-\mathcal{D}^{'}$ can be deduced
as follows. Firstly, relying on the monotonically increasing property of $F_{\min}\left(\cdot\right)$, the minimum value of $F(\mathbf{w})$ over $\mathcal{D}_{1}-\mathcal{D}^{'}$ is $\mathfrak{m}'$. Secondly, recalling that $\mathfrak{m}$ is the minimum value of $F(\mathbf{w})$ over $\mathcal{D}-\mathcal{D}_{1}$ and $\mathfrak{m}'<\mathfrak{m}$, then the minimum value of $F\left(\mathbf{w}\right)$ over $\mathcal{D}-\mathcal{D}'$ is $\mathfrak{m}'$. Therefore, if $F\left(\mathbf{w}\right)<\mathfrak{m}'<\mathfrak{m}$, $\mathbf{w}$ must be in $\mathcal{D}'=\left\{ \mathbf{w}\left|\eu{\mathbf{w}-\mathbf{w}_{0}}<F_{\min}^{-1}\left(\mathfrak{m}'\right)\right.\right\} \cap D$. Otherwise, if a point $\mathbf{\mathbf{w}}_{t}$ satisfying $F\left(\mathbf{w}_t\right)<\mathfrak{m}'$ were in  $\mathcal{D}-\mathcal{D}'$, $F\left(\mathbf{w}_t\right)$
  would be less than the minimum value of $F\left(\mathbf{w}\right)$ over $\mathcal{D}-\mathcal{D}'$, which is a self-contradiction.  Lemma 4 has been proved.
\end{IEEEproof}
Furthermore, since $F_{\min}(r)$ is continuous at $r=0$, the distance given by the
left-hand side of \eqref{eq:invF} can be made arbitrarily small by
selecting a sufficiently small $\mathfrak{m}'$.

Now, let us continue proving Property 2) of Theorem 1. First, note that if
{\small{$\left\Vert \varPi_{Y^{n}}-P_{{U_{1},U_{2}}}(\varPi_{{V_{1}^{n}}\left|{U_{1}^{n}}\right.}^{T}\otimes\varPi_{{V_{2}^{n}}\left|{U_{2}^{n}}\right.}^{T})P_{Y\left|{V_{1},V_{2}}\right.}^{T}\right\Vert _{2} \leq \mu_{1}$}}, with the aid of
triangular inequality, we obtain (\ref{upper_bound}), as given on the next page.
\begin{figure*}[!ht]
\hrulefill
\normalsize
\begin{align}
\nonumber  &
  \Big| \|P_{{U_{1},U_{2}}}(\hat{\varPi}_{{V_{1}^{n}}\left|{U_{1}^{n}}\right.}^{T}\otimes\hat{\varPi}_{{V_{2}^{n}}\left|{U_{2}^{n}}\right.}^{T})P_{Y\left|{V_{1},V_{2}}\right.}^{T}-P_{{U_{1},U_{2}}}P_{Y\left|{V_{1},V_{2}}\right.}^{T}\|_{2}-\|P_{{U_{1},U_{2}}}(\varPi_{{V_{1}^{n}}\left|{U_{1}^{n}}\right.}^{T}\otimes\varPi_{{V_{2}^{n}}\left|{U_{2}^{n}}\right.}^{T})P_{Y\left|{V_{1},V_{2}}\right.}^{T}-P_{{U_{1},U_{2}}}P_{Y\left|{V_{1},V_{2}}\right.}^{T}\|_{2}
\Big| \\
\nonumber& \leq
\left\Vert P_{{U_{1},U_{2}}}(\hat{\varPi}_{{V_{1}^{n}}\left|{U_{1}^{n}}\right.}^{T}\otimes\hat{\varPi}_{{V_{2}^{n}}\left|{U_{2}^{n}}\right.}^{T})P_{Y\left|{V_{1},V_{2}}\right.}^{T}-P_{{U_{1},U_{2}}}(\varPi_{{V_{1}^{n}}\left|{U_{1}^{n}}\right.}^{T}\otimes\varPi_{{V_{2}^{n}}\left|{U_{2}^{n}}\right.}^{T})P_{Y\left|{V_{1},V_{2}}\right.}^{T}\right\Vert _{2}\\
\nonumber& \leq
\left\Vert P_{{U_{1},U_{2}}}(\hat{\varPi}_{{V_{1}^{n}}\left|{U_{1}^{n}}\right.}^{T}\otimes\hat{\varPi}_{{V_{2}^{n}}\left|{U_{2}^{n}}\right.}^{T})P_{Y\left|{V_{1},V_{2}}\right.}^{T}-\varPi_{Y^{n}}\right\Vert _{2}+\left\Vert P_{{U_{1},U_{2}}}(\varPi_{{V_{1}^{n}}\left|{U_{1}^{n}}\right.}^{T}\otimes\varPi_{{V_{2}^{n}}\left|{U_{2}^{n}}\right.}^{T})P_{Y\left|{V_{1},V_{2}}\right.}^{T}-\varPi_{Y^{n}}\right\Vert _{2} \\
& \leq  \mu_1 + \mu.
\label{upper_bound}
\end{align}
\hrulefill
\end{figure*}
Based on $\sum_{m=1}^{2}\|\varPi_{V_{m}^{n}|U_{m}^{n}}-I \|_{2} \leq \delta$, \eqref{upper_bound} implies that
\begin{align}
\nonumber&\|P_{{U_{1},U_{2}}}(\hat{\varPi}_{{V_{1}^{n}}\left|{U_{1}^{n}}\right.}^{T}\otimes\hat{\varPi}_{{V_{2}^{n}}\left|{U_{2}^{n}}\right.}^{T})P_{Y\left|{V_{1},V_{2}}\right.}^{T}-P_{{U_{1},U_{2}}}P_{Y\left|{V_{1},V_{2}}\right.}^{T}\|_{2}\\
&\hspace{100pt}\leq\mu+\mu_{1}+\delta^{2}\left|\mathcal{U}_{1}\right|^{3}\left|\mathcal{U}_{2}\right|^{3}.
\label{eq:ed}
\end{align}The bound on the
right-hand side of \eqref{eq:ed} can be made smaller
than $\mathfrak{m}$ of Lemma~\ref{thm:lem3}, provided that $\delta$, $\mu_1$ and $\mu$ are
all sufficiently small. Thus, applying Lemma~\ref{thm:lem3}, we arrive at
\begin{equation}
D(Y^n)  \leq \left(|\mathcal{U}_{1}|^{2} +|\mathcal{U}_{2}|^{2}\right)  \cdot F^{-1}_{\min}
\left(\mu+\mu_{1}+\delta^{2}\left|\mathcal{U}_{1}\right|^{3}\left|\mathcal{U}_{2}\right|^{3}
\right).
\label{eq:DN}
\end{equation}
Finally,
setting both $\mu_1$ and $\mu$ to $\delta$,
\eqref{eq:DN} proves Property 2) of Theorem 1, where we have {\small{$$\varepsilon(\delta) \defn
\left(|\mathcal{U}_{1}|^{2} +|\mathcal{U}_{2}|^{2}\right) F^{-1}_{\min}\left( 2\delta +
  \delta^{2}\left|\mathcal{U}_{1}\right|^{3}\left|\mathcal{U}_{2}\right|^{3}\right),$$}}which vanishes
as $\delta$ decreases to $0$.

\subsection{Proof of Necessity of Theorem 1}
We assume that the observation channel is manipulable, i.e., there exist
$\Upsilon_{1}$ and $\Upsilon_{2}$ satisfying
\begin{enumerate}
  \item $\overset{2}{\underset{m=1}{\sum}}\left\Vert \Upsilon_{m}-I\right\Vert _{2}^{2}
  >0$,
  \item For an arbitrary value of $m\in\left\{ 1,2\right\}$, $\Upsilon_{m}$ is a stochastic
    matrix, and
  \item $P_{{U_1,U_2}}\left(\overset{2}{\underset{m=1}{\otimes}}\Upsilon_{m}\right)^{T}P_{Y\left|{V_1,V_2}\right.}^{T}=P_{{U_1,U_2}}P_{Y\left|{V_1,V_2}\right.}^{T}$.
\end{enumerate}
Now, let us consider the following two cases:
\begin{enumerate}
  \item[i)] The relay $m$ modifies its input symbols according to $P_{V_{m}|U_{m}} =
 \Upsilon_{m}$, where $m=1,2$, in an
 i.i.d manner. In order to avoid confusion, we use $\varPi'_{V_{m}^{n}|U_{m}^{n}}$ 
 to represent the value of $\varPi_{V_{m}^{n}|U_{m}^{n}}$ obtained in such case, 
 i.e., $\varPi_{V_{m}^{n}|U_{m}^{n}} =
 \varPi'_{V_{m}^{n}|U_{m}^{n}}$. Due to the i.i.d attack, we get $\varPi'_{V_{m}^{n}|U_{m}^{n}}\rightarrow P_{V_{m}|U_{m}}$ for 
 $m=1,2$ according to the law of large numbers. Recall that $P_{V_{m}|U_{m}} =
 \Upsilon_{m}$, we get $\sum_{m=1}^{2} \|\varPi'_{V_{m}^{n}|U_{m}^{n}} - I\|_2 > 0$ and
\begin{equation}
 P_{{U_1,U_2}}\left(\overset{2}{\underset{m=1}{\otimes}}P_{V_{m}|U_{m}} \right)^{T}P_{Y\left|{V_1,V_2}\right.}^{T}=P_{{U_1,U_2}}P_{Y\left|{V_1,V_2}\right.}^{T}.
\label{eq:necc}
\end{equation}

\item[ii)] Neither of the relays is malicious, i.e.,
  $\varPi_{V_{m}^{n}|U_{m}^{n}}=I$ for $m=1,2$.
\end{enumerate}
In both cases, $Y^n$ is an i.i.d sequence, whose distribution
only depends on the distribution of $Y$. In the case i), the distribution of
$Y$ is $P_{{U_1,U_2}}\left(\overset{2}{\underset{m=1}{\otimes}}P_{V_{m}|U_{m}} \right)^{T}P_{Y\left|{V_1,V_2}\right.}^{T}$.
In the case ii), the distribution of $Y$ is $P_{{U_1,U_2}}P_{Y\left|{V_1,V_2}\right.}^{T}$.
From \eqref{eq:necc}, we can see that the distributions of $Y^n$ in
both cases are exactly the same. Thus, any decision statistic $D(Y^n)$
has the same (conditional) distribution in the above-mentioned two cases. Nevertheless,
in the case ii), since $\sum_{m=1}^{2} \|\varPi_{V_{m}^{n}|U_{m}^{n}} - I\|_2 =
0$, Property 2) of Theorem~\ref{thm:main} requires that the probability of
the event $\{ D\left({Y}^{n}\right) > \delta\}$ is arbitrarily small as long as $\delta$
is sufficiently small and $n$ is sufficiently large. On the other hand, in the case
i), by choosing $\delta$ that satisfies $\delta < \sum_{m=1}^{2} \|\varPi'_{V_{m}^{n}|U_{m}^{n}} -
I\|_2$, Property 1) of Theorem~\ref{thm:main} requires that the probability of the same event
$\{D\left({Y}^{n}\right) > \delta\}$ is arbitrarily close to $1$ as long as $n$ is
large enough. Hence, these two requirements lead to contradiction. Therefore, the necessity of Theorem 1
has been proved.

\bibliographystyle{IEEEtran}

\begin{IEEEbiography}[{\includegraphics[width=1in,height=1.25in,clip,keepaspectratio]{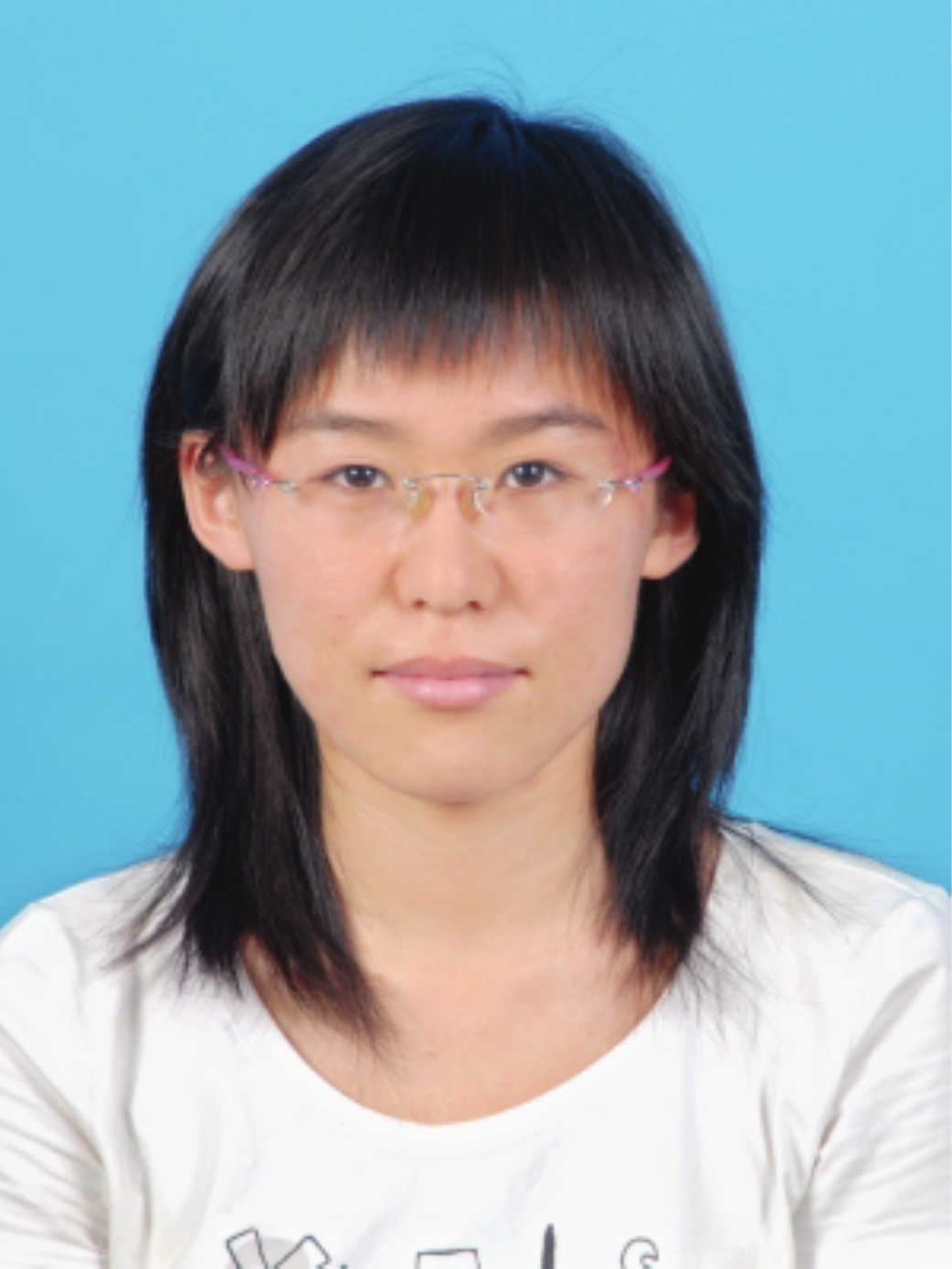}}]{Ruohan Cao} received her B.Eng. degree in 2009 from Shandong University of Science and Technology (SDUST), Qingdao, China. She received the Ph.D. degree in 2014
form Beijing University of Posts and Telecommunications (BUPT), Beijing, China. From November 2012 to August 2014, she also served as a research assistant for the Department of Electrical and Computer Engineering at University of Florida, supported by the China Scholarship Council. 
She is now with the Institute of Information Photonics and Optical Communications, BUPT, as a Postdoc. 
Her research interests include physical-layer network coding, multiuser multiple-input-multiple-output systems and physical-layer security.
\end{IEEEbiography}

\begin{IEEEbiography}[{\includegraphics[width=1in,height=1.25in,clip,keepaspectratio]{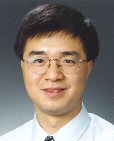}}]{Tan F. Wong} received the B.Sc. degree (first class honors) in electronic engineering
from the Chinese University of Hong Kong in 1991 and the M.S.E.E. and Ph.D. degrees
in electrical engineering from Purdue University in1992 and 1997, respectively.

He was a Research Engineer at the Department of Electronics, Macquarie University,
Sydney, Australia. He also served as a Postdoctoral Research Associate at the School
of Electrical and ComputerEngineering, Purdue University. Since August 1998, he has
been with the University of Florida, where he is currently a Professor of Electrical and
Computer Engineering. Dr. Wong was the Editor for Wideband and Multiple Access Wireless
Systems for the IEEE TRANSACTIONS ON COMMUNICATIONS and was the
Editor-in-Chief for the IEEE TRANSACTIONS ON VEHICULAR TECHNOLOGY. He also
served as an Associate Editor for the IEEE SIGNAL PROCESSING LETTERS.
\end{IEEEbiography}

\begin{IEEEbiography}[{\includegraphics[width=1in,height=1.25in,clip,keepaspectratio]{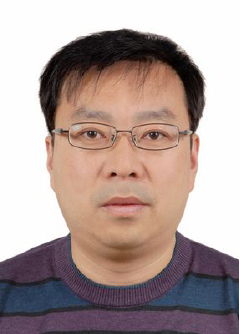}}]{Tiejun Lv }(M'08-SM'12) received the M.S. and Ph.D. degrees in electronic engineering from the University of Electronic Science and Technology of China (UESTC), Chengdu, China, in 1997 and 2000, respectively. From January 2001 to January 2003, he was a Postdoctoral Fellow with Tsinghua University, Beijing, China. In 2005, he became a Full Professor with the School of Information and Communication Engineering, Beijing University of Posts and Telecommunications (BUPT). From September 2008 to March 2009, he was a Visiting Professor with the Department of Electrical Engineering, Stanford University, Stanford, CA, USA. He is the author of more than 200 published technical papers on the physical layer of wireless mobile communications. His current research interests include signal processing, communications theory and networking. Dr. Lv is also a Senior Member of the Chinese Electronics Association. He was the recipient of the Program for New Century Excellent Talents in University Award from the Ministry of Education, China, in 2006. He received the Nature Science Award in the Ministry of Education of China for the hierarchical cooperative communication theory and technologies in 2015.
\end{IEEEbiography}

\begin{IEEEbiography}[{\includegraphics[width=1in,height=1.25in,clip,keepaspectratio]{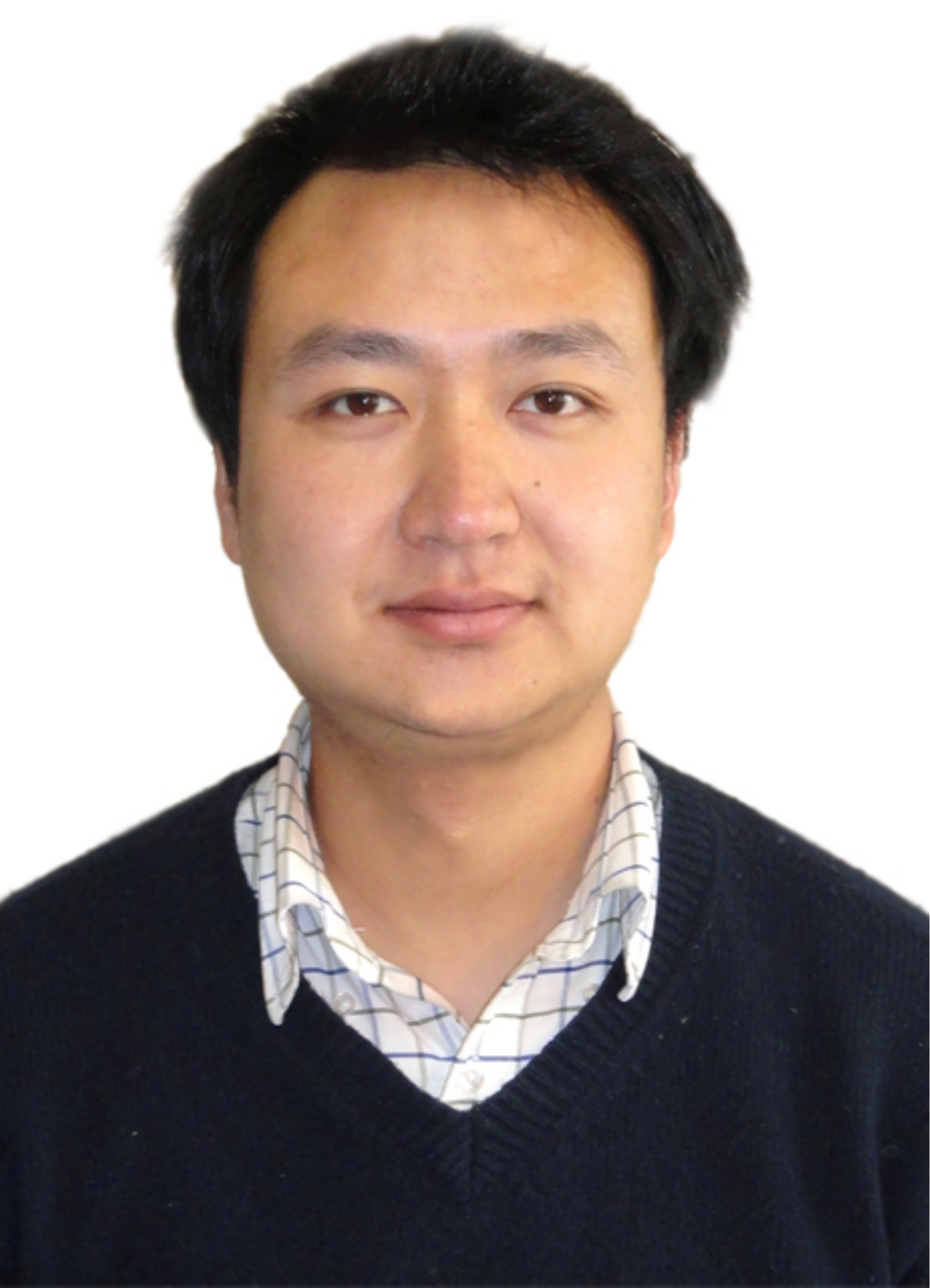}}]{Hui Gao} S'10-M'13-SM'16) received his B. Eng. degree in Information Engineering and Ph.D. degree in Signal and Information Processing from Beijing University of Posts and Telecommunications (BUPT), Beijing, China, in July 2007 and July 2012, respectively. From May 2009 to June 2012, he also served as a research assistant for the Wireless and Mobile Communications Technology R$\&$D Center, Tsinghua University, Beijing, China. From Apr. 2012 to June 2012, he visited Singapore University of Technology and Design (SUTD), Singapore, as a research assistant. From July 2012 to Feb. 2014, he was a Postdoc Researcher with SUTD. He is now with the School of Information and Communication Engineering, Beijing University of Posts and Telecommunications (BUPT), as an assistant professor. His research interests include massive MIMO systems, cooperative communications, ultra-wideband wireless communications.
\end{IEEEbiography}

\begin{IEEEbiography}[{\includegraphics[width=1in,height=1.25in,clip,keepaspectratio]{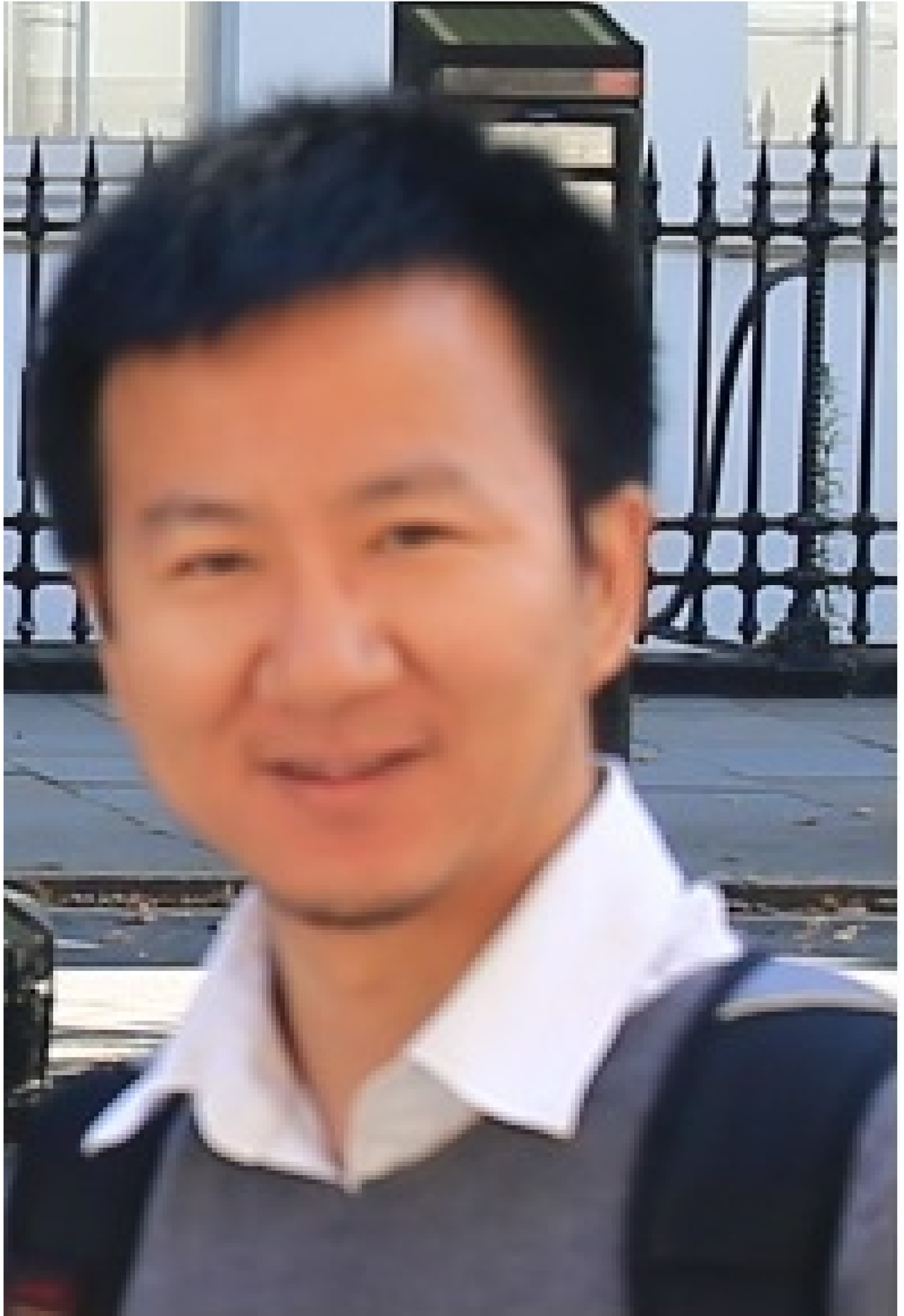}}]{Shaoshi Yang} (S'09-M'13) received his B.Eng. degree in Information Engineering from Beijing University of Posts and Telecommunications (BUPT), 
Beijing, China in Jul. 2006, his first Ph.D. degree in Electronics and Electrical Engineering from University of Southampton, U.K. 
in Dec. 2013, and his second Ph.D. degree in Signal and Information Processing from BUPT in Mar. 2014. He is now working as 
a Postdoctoral Research Fellow in University of Southampton, U.K. From November 2008 to February 2009, he was an Intern Research 
Fellow with the Communications Technology Lab (CTL), Intel Labs, Beijing, China, where he focused on Channel Quality Indicator 
Channel (CQICH) design for mobile WiMAX (802.16m) standard. His research interests include MIMO signal processing, green radio, 
heterogeneous networks, cross-layer interference management, convex optimization and its applications. He has published 
in excess of 30 research papers on IEEE journals. 

Shaoshi has received a number of academic and research awards, including the prestigious Dean's Award for Early Career Research 
Excellence at University of Southampton, the PMC-Sierra Telecommunications Technology Paper Award at BUPT, the Electronics and 
Computer Science (ECS) Scholarship of University of Southampton, and the Best PhD Thesis Award of BUPT. He is a member of IEEE/IET, 
and a junior member of Isaac Newton Institute for Mathematical Sciences, Cambridge University, U.K. He also serves as a TPC member 
of several major IEEE conferences, including \textit{IEEE ICC, GLOBECOM, VTC, WCNC, PIMRC, ICCVE, HPCC}, and as a Guest Associate 
Editor of \textit{IEEE Journal on Selected Areas in Communications.} ({http://sites.google.com/site/shaoshiyang/}) \end{IEEEbiography}

 \end{document}